\documentclass{article}
\usepackage{graphicx} 

\usepackage[utf8]{inputenc}
\usepackage{amstext,amsmath,amsthm,verbatim,amssymb,amsfonts,amscd, graphicx}
\usepackage{graphics,caption}
\usepackage{hyperref}
\hypersetup{
    colorlinks=true,   	
    linkcolor=red,      
    citecolor = [rgb]{0 0.7 0},   	
    filecolor=magenta, 	
    urlcolor=blue
}
\usepackage{enumitem}
\usepackage{apptools}
\usepackage{titlesec}
\usepackage{natbib,comment}
\usepackage{color,bm}
\usepackage[baseline]{euflag}

\title{Entropic additive energy
and entropy inequalities\\ for sums and products} 

\author{Rupert Li\thanks{DPMMS,
	University of Cambridge,
	Centre for Mathematical Sciences,
        Wilberforce Road,
	Cambridge CB3 0WB, U.K.
        Email: {\tt rupertli@stanford.edu}.
	}
\and
Lampros Gavalakis%
	\thanks{Statistical Laboratory, DPMMS,
	University of Cambridge,
	Centre for Mathematical Sciences,
        Wilberforce Road,
	Cambridge CB3 0WB, U.K.
        Email: {\tt lg560@cam.ac.uk}
	}
\and 
Ioannis Kontoyiannis%
	\thanks{Statistical Laboratory, DPMMS,
	University of Cambridge,
	Centre for Mathematical Sciences,
        Wilberforce Road,
	Cambridge CB3 0WB, U.K.
        Email: {\tt yiannis@maths.cam.ac.uk}.
	I.K.\ was supported in part
	by the EPSRC funded INFORMED-AI project EP/Y028732/1.
}
}
\date{\today}

\newcommand{\set}[1]{\left\{#1\right\}}
\newcommand{\abs}[1]{\left\lvert#1\right\rvert}

\newcommand{\paren}[1]{\left(#1\right)}

\newcommand{\RL}{{\mathbb R}}

\newcommand{\IN}{{\mathbb Z}}

\newcommand{\IND}{{\mathbb I}}
\newcommand{\BBP}{{\mathbb P}}

\newcommand{\BBE}{{\mathbb E}}

\def\ba{\begin{align}}
\def\ea{\end{align}}
\def\ban{\begin{align*}}
\def\ean{\end{align*}}

\def\be{\begin{eqnarray}}
\def\ee{\end{eqnarray}}
\def\ben{\begin{eqnarray*}}
\def\een{\end{eqnarray*}}

\def\bqq{\begin{equation}}
\def\eqq{\end{equation}}
\def\bqqn{\begin{equation*}}
\def\eqqn{\end{equation*}}

\def\sq{$\Box$}

\def\qed{\ifmmode\sq\else{\unskip\nobreak\hfil
\penalty50\hskip1em\null\nobreak\hfil\sq
\parfillskip=0pt\finalhyphendemerits=0\endgraf}\fi\par\medbreak}

\newsavebox{\junk}
\savebox{\junk}[1.6mm]{\hbox{$|\!|\!|$}}

\def\til={{\widetilde =}}

 \def\eq#1/{(\ref{#1})}

\newtheorem{theorem}{Theorem}[section]
\newtheorem{corollary}[theorem]{Corollary}
\newtheorem{proposition}[theorem]{Proposition}
\newtheorem{lemma}[theorem]{Lemma}

\newtheorem{example}[theorem]{Example}

\def\eq#1/{(\ref{e:#1})}

\def\bdes{\begin{description}}
\def\edes{\end{description}}

\def\notes#1{}

\definecolor{mag}{rgb}{0.7,0,0.3}
\definecolor{dgreen}{rgb}{0.1,0.5,0.1}
\definecolor{dred}{rgb}{.8,0,0}
\definecolor{gray}{rgb}{.8,.8,.8}
\definecolor{brown}{rgb}{0.6451,0.3706,0.1745}

\begin{document}

\maketitle

\begin{abstract}
Following a growing number of studies 
that, over the past 15 years, have 
established entropy inequalities via ideas and tools
from additive combinatorics, in this work
we obtain a number of new bounds for the 
differential entropy of sums, products, and 
sum-product combinations
of continuous random variables. 
Partly motivated by recent work by Goh on
the discrete entropic version of the notion
of ``additive energy'', we introduce
the additive energy of pairs of continuous 
random variables and prove
various versions of the
statement that ``the additive
energy is large if and only if the entropy
of the sum is small'', along with 
a version of the
Balog--Szemer\'{e}di--Gowers theorem for differential entropy.
Then, motivated in part by recent work 
by M\'ath\'e and O'Regan,
we establish a series of new differential
entropy inequalities for products 
and sum-product combinations
of continuous random variables.
In particular, we prove a new,
general, ring Pl\"unnecke--Ruzsa entropy inequality.
We briefly return to the case of discrete
entropy and provide a characterization of discrete
random variables with ``large doubling'', 
analogous to Tao's Freiman-type inverse
sumset theory for the case of small doubling.
Finally, we consider the natural entropic
analog of the 
Erd\H{o}s--Szemer\'{e}di sum-product phenomenon
for integer-valued random variables. 
We show that, if it does hold, then 
the range of parameters for which it does 
would necessarily be significantly more
restricted than its anticipated
combinatorial counterpart.
\end{abstract}

\section{Introduction}
\label{section:introduction}

\subsection{Additive combinatorics and entropy inequalities}

The field of {\em additive combinatorics}~\cite{TV06:book} 
studies occurrences 
of additive structures, such as arithmetic progressions,
in subsets of abelian groups like the additive group
of the integers $(\IN,+)$.
The Pl\"unnecke--Ruzsa \emph{sumset theory} is an important
sub-field of additive combinatorics that studies
\emph{sumset inequalities}, which are used
in many branches of combinatorics as well as
in other areas of mathematics.

The \emph{sumset} $A+B$ of two discrete subsets $A$ and $B$ of an 
additive group $G$, i.e., an abelian group $(G,+)$, is 
$A+B=\set{a+b:a\in A,b\in B}$. Sumset inequalities 
include the trivial bound
\begin{equation}
\abs{A+B} \geq \max\set{\abs{A},\abs{B}},
\label{old_eq:trivial_sumset_bounds}
\end{equation}
as well as 
many more nontrivial and subtle results.
For example,
if we similarly define the \emph{difference set} 
$A-B=\set{a-b:a\in A,b\in B}$,
then the {\em Ruzsa triangle inequality}~\cite{ruzsa:78} 
says that 
the cardinality of $A-C$ 
can be bounded in terms of the cardinalities of
$A-B$ and $B-C$ 
via
\begin{equation}
    \abs{A-C} \leq \frac{\abs{A-B}\abs{B-C}}{\abs{B}},
\label{old_eq:Rusza_triangle_sumset}
\end{equation}
and the \emph{sum-difference inequality}~\cite{ruzsa:96} 
compares the cardinality of $A+B$ with that of $A-B$:
\bqq
\abs{A+B} \leq \frac{\abs{A-B}^3}{\abs{A}\abs{B}}.
\label{eq:sumdiffcomb}
\eqq

Shannon's Asymptotic Equipartition
Property (AEP) motivates an interesting connection
between sumset inequalities and information
theory.
Let $X$ be a discrete random variable with
values in a set
$A$, and with entropy $H=H(X)$, in nats.
Suppose $X_1,X_2,\ldots$ are independent
and identically distributed (i.i.d.)\ copies of $X$.
Then
the AEP tells us that
there is a subset $B^*_n$ of $A^n$ such that,
for large $n$,
$B_n^*$ carries
essentially all of the probability of the distribution
of $(X_1,\ldots,X_n)$, i.e., 
$\BBP((X_1,\ldots,X_n)\in B^*_n)= 1-o(1)$,
and its cardinality is 
minimal: It contains $|B^*_n|= e^{n(H+o(1))}$ 
elements, while any other subset $B_n$ of $A^n$
with
$\BBP((X_1,\ldots,X_n)\in B_n)= 1-o(1)$ is 
necessarily at least as large,
$|B_n|\geq e^{n(H-o(1))}.$
Therefore, the random variables
$(X_1,\ldots,X_n)$ are essentially 
supported on a set of size $\doteq e^{nH}$,
which is typically much smaller than
the size $|A|^n=e^{n\log |A|}$ of all 
of $A^n$. In this sense, we may think
of $e^H$ as the cardinality of the ``essential
support'' of $X$, and consequently we can
interpret the entropy $H(X)$ of a discrete
random variable $X$ as a ``probabilistic analog''
of the log-cardinality of a discrete set
$A$.

Thus motivated, 
Ruzsa~\cite{Ruz09} in 2009 
and Tao~\cite{tao:10} in 2010 
began a systematic exploration of this
entropy/cardinality
correspondence, in the context of sumset 
inequalities. Specifically, they asked
whether, starting with an arbitrary sumset bound 
and replacing discrete subsets of an additive group
with independent discrete random 
variables,
and log-cardinalities by entropies,
would lead to a legitimate new entropy
inequality. Although far from obvious, 
the answer generally turns out to be 
``yes''. 
For example, for independent random variables $X$ and $Y$,
the simple bound in~\eqref{old_eq:trivial_sumset_bounds}
corresponds to the familiar, elementary entropy inequality
\begin{equation*}
    H(X+Y)\geq \max\set{H(X),H(Y)}.
\end{equation*}
Similarly~\cite{tao:10}, for independent $X,Y,Z$, the entropic version
of the Ruzsa triangle
inequality in~(\ref{old_eq:Rusza_triangle_sumset})
is
$$H(X-Z)+H(Y)\leq H(X-Y)+H(Y-Z),$$
and the sum-difference inequality 
in~(\ref{eq:sumdiffcomb}) becomes
$$H(X+Y)+H(X)+H(Y)\leq 3H(X-Y).$$
This correspondence 
is rich and nontrivial, with neither the combinatorial
nor the entropic version of these bounds being
formally stronger than the other. 

The entropy inequalities resulting from the
entropy/cardinality correspondence
have been extended and generalized along
several directions,
e.g., in~\cite{KM:16,makkuva:18,nair:23,gavalakis-doubling:24,goh:24,green:25},
and they have
found numerous applications in core information-theoretic
problems; see, 
e.g.,~\cite{lapidoth-pete,stotz:16,stotz:22,nair:23ITT,yao:24}.
In the reverse direction,
entropic sumset bounds have also been finding important
applications to questions in combinatorics. For example, 
the recent resolution by Gowers, Green, Manners and 
Tao~\cite{GGMT:23b,GGMT:24} 
of two important cases of the \emph{polynomial Freiman--Ruzsa conjecture} 
critically depends on the entropic formulation of this
purely combinatorial problem. 

\subsection{Differential entropy inequalities}

Given the wealth and utility of the discrete entropy
inequalities produced 
via the entropy/cardinality correspondence, it is natural
to ask whether they extend to the case of the differential
entropy for continuous random variables, {i.e.,
for real-valued random variables with densities 
with respect to Lebesgue measure on~$\RL$}. Additional motivation
for this comes from the continuous version of the AEP:
Similarly to the discrete case, the differential entropy
$h=h(X)$ of a continuous random variable $X$ may be interpreted
as the size (in the sense of Lebesgue measure) of the
essential support of $X$. As it turns out,
it is generally possible to 
translate discrete entropic sumset inequalities to 
corresponding differential entropy bounds,
but it is not entirely straightforward.

An essential element in the proofs of the 
discrete entropic
versions of most sumset inequalities is
the following observation:
Consider two arbitrary discrete random variables
$X,Y$. Suppose that the random variable
$Z$ is a function of the pair $(X,Y)$, 
i.e., $Z=\phi(X,Y)$ for some function $\phi$,
and that the random variable $W$ can
be expressed as a function of either $X$ or $Y$,
i.e., $W=f(X)=g(Y)$ for appropriate functions
$f,g$. Then it is not hard to
show~\cite{tao:10} that
\[ H(W) + H(Z) \leq H(X) + H(Y). \]
This property is sometimes
referred to as the {\em functional submodularity}
of Shannon entropy.

It is easy to come up with examples showing that functional 
submodularity fails in the case of differential entropy~\cite{KM:14}.
Moreover,
the steps in which functional submodularity is used
in the original proofs of discrete entropic sumset bounds involve
intermediate inequalities that are not valid for continuous
random variables and differential entropy.
Nevertheless, 
as was shown in~\cite{KM:14}
and~\cite{KM:16}, essentially all of the earlier
discrete entropic results obtained by
Ruzsa~\cite{Ruz09}, Tao~\cite{tao:10},
and others have near-identical continuous analogs,
but instead of functional submodularity, the key
ingredient in their proofs is the data processing
property of mutual information.

Indeed, for any independent continuous random variables
$X,Y,Z$, we have the elementary bound
$$h(X+Y)\geq \max\{h(X),h(Y)\},$$
the Ruzsa triangle inequality~\cite{KM:14}
becomes
$$h(X-Z)+h(Y)\leq h(X-Y)+h(Y-Z),$$
and the continuous analog of the
sum-difference inequality~\cite{KM:14} is simply
$$h(X+Y)+h(X)+h(Y)\leq 3h(X+Y).$$

In this paper, we continue the exploration
of continuous versions of discrete entropy sumset bounds,
largely motivated by recent results obtained
by Goh~\cite{goh:24} and by M\'ath\'e and 
O'Regan~\cite{mathe:23}. In the process,
we also examine some questions regarding
the entropy of sums and products
of discrete random variables.
As in the earlier work~\cite{KM:14,KM:16},
the main obstacle to obtaining continuous
generalizations of the discrete bounds in~\cite{goh:24}
and~\cite{mathe:23} is the fact that their proofs
rely heavily on the functional submodularity
property of discrete entropy.
Therefore, substantially new arguments
are required, and the
data processing property of mutual information
again plays a key role.

\subsection{Outline of main results}

After a preliminary discussion of earlier
work in Section~\ref{section:preliminaries},
our main results, briefly outlined below,
are developed in
Sections~\ref{section:entropic_additive_energy}%
--\ref{section:entropic_sumproduct}.

\medskip

\noindent
{\bf Additive energy and the differential entropy of sums.}
In Section~\ref{section:entropic_additive_energy} we define
the {\em additive energy}, $a(X,Y)$, for pairs of continuous
random variables $(X,Y)$, and we observe that it can be 
equivalently
expressed as $a(X,Y)=2h(X,Y)-h(X+Y)$.
This representation immediately leads to the 
main subject of 
this section, which is to establish various 
quantitative versions of the
statement that ``if $a(X,Y)$ is large (close to its upper bound) then $h(X+Y)$
is small (close to its lower bound)'', as well as its converse. In this sense, 
the additive energy $a(X,Y)$ can naturally be interpreted
as a measure of the ``additive structure'' present
in $X$ and $Y$, in close analogy with its discrete counterpart
examined in~\cite{goh:24}.

\medskip

\noindent
{\bf Balog--Szemer\'{e}di--Gowers theorem for differential entropy.}
In Section~\ref{section:BSG}, we prove 
a version of the Balog--Szemer\'edi--Gowers (BSG) theorem 
for differential entropy and discuss its relation
with earlier (both discrete and continuous) versions.
Roughly speaking, the BSG theorem is the important
assertion that, if the additive energy $a(X,Y)$ is 
appropriately ``large'', then we can find a conditioning
random variable $Z$ such that $X$ and $Y$ are 
approximately conditionally independent given $Z$,
the values of the entropies $h(X|Z)$ and $h(Y|Z)$ 
are close to those of the original entropies 
$h(X)$ and $h(Y)$, respectively, and $h(X+Y|Z)$ is
appropriately small.

\medskip

\noindent
{\bf Stability of large discrete entropic doubling. }
In Section~\ref{section:Sidon_stability}
we briefly return to the discrete setting,
and we consider the problem of characterizing
discrete random variables $X$ with the property
that, when $X$ and $X'$ are i.i.d., 
then the entropy $H(X+X')$ of their sum is close
to its maximum possible value, $2H(X)$. 
This complements the Freiman-type inverse
sumset theory 
of Tao for the case when $H(X+X')$ is close
to its obvious lower bound, $H(X)$. 
Interestingly, we find that, when 
$H(X+X')$ is appropriately large, then
$X$ is necessarily approximately supported
on a {\em Sidon set}, i.e., a set $A$ for which
the sums $a+b$ of all pairs of elements $a,b\in A$
are unique.

\medskip

\noindent
{\bf Differential entropy bounds for random products.}
In Section~\ref{section:product}, we discuss how many 
of the entropic sumset inequalities in earlier sections
and in earlier work can be generalized for {\em products}
rather than sums of (both discrete and continuous) random 
variables. The results in this section mostly follow directly
from the general development in~\cite{KM:16}, 
and they provide useful tools for the new inequalities
developed next. 

\medskip

\noindent
{\bf Sum-product inequalities for differential entropy.}
The previous section provides important tools 
for Section~\ref{section:sum_product}, which 
extends recent work of M\'ath\'e and O'Regan~\cite{mathe:23} 
on entropic {\em sum-product} entropic inequalities,
namely, entropy bounds that involve both sums and products
of random variables. The main result in this section is 
a new and general ``ring Pl\"unnecke--Ruzsa entropy inequality''.
It gives an upper bound on the entropy of sums of products
of i.i.d.\ random variables, in terms of their marginal
entropy and their associated pairwise ``doubling'' constants,
introduced and discussed in Section~\ref{s:multisX}.

\medskip

\noindent
{\bf On the entropic Erd\H{o}s--Szemer\'{e}di sum-product phenomenon.}
Finally, in Section~\ref{section:entropic_sumproduct} we
ask whether 
an entropic Erd\H{o}s--Szemer\'{e}di sum-product phenomenon
exists for integer-valued random variables.
Specifically, we ask whether it is necessarily the case
that, if $X$ and $X'$ are i.i.d.\ with values in $\IN$, 
then at least one of $H(X+X')$ and $H(XX')$ must be
significantly larger than $H(X)$. We do not present any 
positive results, but we 
show, via example, that the range of the validity of
the naive version of this conjecture must necessarily
be more restricted than both what is known and what is
conjectured to hold in the combinatorial case.

\section{Preliminaries and some prior work}
\label{section:preliminaries}

\subsection{Combinatorial background}
\label{subsec:entropy_methods}

In this section we describe the main 
concepts and results from 
additive combinatorics
that motivate our subsequent entropic bounds.
In particular, we introduce the combinatorial versions
of the Ruzsa distance and the additive energy, 
and we describe those of their properties that will be relevant
to the development of our results in 
Sections~\ref{section:entropic_additive_energy}%
--\ref{section:entropic_sumproduct}.
All of the results in this section
can be found, e.g., in the text~\cite{TV06:book}.

Let $G$ be an additive group, that is, an arbitrary 
abelian group $(G,+)$. The
\emph{Ruzsa distance} $d(A,B)$ between two finite
subsets $A$ and $B$ of $G$ is
\begin{equation}\label{old_eq:Ruzsa distance}
    d(A,B)=\log\frac{\abs{A-B}}{\sqrt{\abs{A}\abs{B}}},
\end{equation}
where, throughout the paper, `$\log$' denotes the natural logarithm.
Note that the Ruzsa distance is not a metric, 
as $d(A,A)\neq0$ in general.
In this notation, the
Ruzsa triangle inequality in~(\ref{old_eq:Rusza_triangle_sumset})
 becomes
\begin{equation}
\label{old_eq:Rusza_triangle}
    d(A,C) \leq d(A,B) + d(B,C).
\end{equation}
Similarly, the sum-difference inequality 
in~(\ref{eq:sumdiffcomb}) can be expressed as
\bqq
d(A,-B)\leq 3d(A,B).
\label{old_eq:sumdiff}
\eqq
In the case $A=B$, the following tighter bounds
are available, sometimes referred to as the
{\em doubling-difference} inequality
\bqq
\frac{1}{2}d(A,A)\leq d(A,-A)\leq 2d(A,A).
\label{old_eq:doublingdiff}
\eqq

The {\em doubling constant} of a finite set $A\subset G$ is
simply the ratio:
\bqq
s(A)=\frac{|A+A|}{|A|}.
\label{eq:doublingA}
\eqq
More generally, for the $k$-fold addition of a 
set $A$ to itself, 
we write
\[ kA
=\set{a_1+a_2+\cdots+a_k:a_1,a_2,\dots,a_k\in A}, \]
for any nonnegative integer $k$.
The Pl\"unnecke--Ruzsa inequality \cite{plunnecke:70,ruzsa:89}
is a fundamental result in additive combinatorics,
which states the following.
For any finite subsets $A,B$ of $G$,
if $\abs{A+B}\leq K\abs{A}$ for some $K>0$, then
for all nonnegative integers $m,n$ we have:
\bqq
\abs{nB-mB}\leq K^{n+m}\abs{A}.
\label{eq:PRsets}
\eqq
This bound is often used when $A=B$, where the best 
constant $K$ is the doubling constant $s(A)$.

The \emph{additive energy} of two subsets $A,B$ of $G$
is
\bqq
E(A,B) 
= 
	\abs{\{(a_1,a_2,b_1,b_2)\in A\times A \times B\times B
	: a_1+b_1=a_2+b_2)\}}
	\label{old_eq:additive_energy}.
\eqq
{
Although it is not an ``energy'' function in the sense of physics,
the natural interpretation of $E(A,B)$ is as a measure of the 
degree of additive structure present in $A+B$;
see, e.g.,~\cite{TV06:book} for extensive discussion of
its properties.} For example, 
$E(A,A)$ is always bounded above by $|A|^3$,
and it is close to $|A|^3$ if and only if $A$
is close to an arithmetic progression.

An elementary but important inequality
for $E(A,B)$ is:
\begin{equation}\label{eq:additive_combinatorics_energy_inequality}
    E(A,B) \geq \frac{\abs{A}^2\abs{B}^2}{\abs{A+B}}.
\end{equation}
One application of this inequality is in showing that
a small sumset $A+B$ implies large additive energy $E(A,B)$,
further reinforcing the interpretation $E(A,B)$ described above.
To see this, note that since 
$\abs{A+B}\geq\max\set{\abs{A},\abs{B}}$, 
we always have $\abs{A+B}\geq\abs{A}^{1/2}\abs{B}^{1/2}$.
We can thus consider $\abs{A+B}$ to be ``small'' 
if $\abs{A+B}\leq C\abs{A}^{1/2}\abs{B}^{1/2}$ for some constant $C>1$.
Then~\eqref{eq:additive_combinatorics_energy_inequality} 
leads to the implication:
\begin{equation}
\label{old_eq:additive_energy_symmetrized}
\abs{A+B}\leq C\abs{A}^{1/2}\abs{B}^{1/2}
\quad
\Rightarrow
\quad
    E(A,B) \geq \frac{1}{C}\abs{A}^{3/2}\abs{B}^{3/2}.
\end{equation}

\subsection{Discrete entropy, Ruzsa distance and additive energy}

Here we describe the entropic analogs of the Ruzsa distance
and the additive energy, together with some of the main 
properties, along the lines of the entropy/cardinality
correspondence. The results in this section
can be found in~\cite{tao:10} and~\cite{goh:24}.
The entropy of an arbitrary discrete random variable
$X$ with probability mass function $P$ on a discrete
alphabet $A$ is defined, as usual, as
$H(X)=\BBE[-\log P(X)]$. In all our results
throughout the paper,
it is implicitly assumed that all entropies that
appear in the statements are finite.

Let $X,Y$ be two discrete random variables with
values in an additive group $G$.
In analogy with~\eqref{old_eq:Ruzsa distance},
the entropic Ruzsa distance $d(X,Y)$ between 
$X$ and $Y$ is defined as
\bqq
d(X,Y) = H(X'-Y')-\frac{1}{2}H(X)-\frac{1}{2}H(Y),
\label{eq:discreteRd}
\eqq
where $X'$ and $Y'$ are independent copies of $X$ and $Y$, respectively.
Once again, we note that this is not a metric,
as $d(X,X)\neq 0$ in general. Also we observe
that $d(X,Y)$ does not depend on the joint distribution
of $(X,Y)$, only on their marginals.

In this notation, the entropic version of the 
Ruzsa triangle inequality~(\ref{old_eq:Rusza_triangle})
states that,
for any three discrete random variables $X,Y,Z$,
\bqq
d(X,Z) \leq d(X,Y) + d(Y,Z),
\label{eq:RuzsaTIH}
\eqq
the entropic version of the sum-difference 
inequality~(\ref{old_eq:sumdiff})
is
\bqq
d(X,-Y)\leq 3d(X,Y),
\label{eq:SDIH}
\eqq
and the entropic doubling-difference inequality
corresponding to~(\ref{old_eq:doublingdiff}) is
\bqq
\frac{1}{2}d(X,X)\leq d(X,-X)\leq 2d(X,X).
\label{eq:DDIH}
\eqq

An entropic analog of the 
Pl\"unnecke--Ruzsa inequality~(\ref{eq:PRsets})
was proved by Tao~\cite{tao:10}.
It says that, if 
$X_1,X_2,\dots,X_n,X_1',X_2',\dots,X_m'$ are independent 
copies of a discrete random variable $X$ taking values in an 
additive group, and $H(X_1+X_1')\leq H(X)+\log K$ for some 
$K\geq1$, then,
\bqq
H(X_1+X_2+\cdots+X_n-X_1'-X_2'-\cdots-X_m')\leq H(X)+O(n+m)\log K.
\label{eq:HPR}
\eqq
This is formally weaker than~(\ref{eq:PRsets}) 
because it loses an absolute constant with respect 
to the exponent $n+m$, and also because the constant 
$K$ in the bound requires 
$X$ and $X'$ to have the same distribution.

Continuing in the spirit of the entropy/cardinality
correspondence, 
Goh~\cite{goh:24} recently introduced a version
of the additive energy of pairs of discrete random 
variables, and explored its properties. 
In analogy with the combinatorial version
of additive energy~(\ref{old_eq:additive_energy}),
the \emph{entropic additive energy} of two jointly 
distributed discrete random variables $X$ and $Y$ 
taking values in the same additive group is
\bqq
A(X,Y)=H(X_1,Y_1,X_2,Y_2,X+Y),
\label{eq:AXY}
\eqq
where $(X_1,Y_1)$ and $(X_2,Y_2)$ are conditionally independent 
versions of $(X,Y)$ given $X+Y$. Intuitively, we can think
of first generating $(X,Y)$ and then independently 
generating $(X_1,Y_1)$ and $(X_2,Y_2)$ from the same distribution
as $(X,Y)$, conditioned on the event $\{X_1+Y_1=X_2+Y_2=X+Y\}$.

We will also find it convenient to use the following
equivalent expression for $A(X,Y)$:
\begin{align}
    A(X,Y)
    	&= H(X_1,Y_1,X_2,Y_2,X+Y)
    \nonumber\\ 
	&= H(X+Y)+H(X_1,Y_1|X+Y)+H(X_2,Y_2|X+Y)
    \nonumber\\ 
	&= 2H(X,Y|X+Y)+H(X+Y)
	\nonumber\\
	&= 2H(X,Y)-H(X+Y).
    	\label{eq:AXY2} 
\end{align}
This expression clearly leads to the same interpretation
as for the combinatorial additive energy:
Small entropy of the sum $X+Y$ implies large
additive energy $A(X,Y)$. This interpretation
is further reinforced by the entropic
analog of the implication~(\ref{old_eq:additive_energy_symmetrized}).
Here we actually have that, if $X,Y$ are independent,
then
\begin{equation}
\label{eq:AXYequivalent}
H(X+Y)\leq \frac{1}{2}H(X)+\frac{1}{2}H(Y)+\log C
\quad
\Leftrightarrow
\quad
    A(X,Y) \geq \frac{3}{2}H(X)+\frac{3}{2}H(Y)-\log C.
\end{equation}
Moreover, for arbitrary pairs of (not necessarily independent)
discrete random variables $(X,Y)$, 
the reverse implication remains valid: If the
additive energy $A(X,Y)$ is large, then 
$H(X+Y)$ is small.

{
Both in the combinatorial case and in the entropy case,
the terms ``large'' and ``small''
are understood in the sense that the relevant constants $C$ are fixed
while the set cardinalities and the random variable entropies may be
arbitrarily large.}

\subsection{Differential entropy and entropic Ruzsa distance}
\label{s:diffh}

As was shown in~\cite{KM:14}, all of the
entropy bounds of the previous section
extend to continuous random variables.
Apart from the proof
of the sum-difference inequality which is new,
all other results in this section can be found in~\cite{KM:14}.
The \emph{differential entropy} $h(X)$ of a continuous random 
vector $X$ with density $f$ in $\RL^d$, or, equivalently,
of a random vector $X=(X_1,\ldots,X_d)$
in $\RL^d$ with joint density $f$, is 
\[ h(X) = -\int_{\RL^d} f(x)\log f(x)\,dx = \BBE[-\log f(X)], \]
whenever the integral exists.
Otherwise, we let $h(X)=-\infty$.
Throughout the paper, the differential entropy of any 
continuous random variable or random vector appearing in the statement 
of any of our results is assumed to exist and be finite.

Similarly to the discrete case~(\ref{eq:discreteRd}), 
the Ruzsa distance between two 
continuous random variables $X,Y$ is
\bqq
d(X,Y) = h(X'-Y')-\frac{1}{2}h(X)-\frac{1}{2}h(Y), 
\label{eq:contRuzsa}
\eqq
where $X'$ and $Y'$ are independent copies of $X$ and $Y$, respectively.
Then, for arbitrary continuous random variables
$X,Y,Z$, the Ruzsa triangle inequality~(\ref{eq:RuzsaTIH}),
the sum-difference inequality~(\ref{eq:SDIH}),
and the doubling-difference inequality~(\ref{eq:DDIH})
remain valid exactly as before.
Another useful bound in the same spirit  is the
{\em submodularity for sums} inequality, which says that,
if $X,Y,Z$ are independent random vectors, then
\begin{equation}
\label{old_eq:sum_submodularity}
    h(X+Y+Z) + h(Y) \leq h(X+Y) + h(Y+Z).
\end{equation}
The obvious analog of~(\ref{old_eq:sum_submodularity})
is also true in the case of discrete entropy~\cite{KV:83}.

The proof 
of the sum-difference inequality for differential
entropy 
in~\cite{KM:14} 
used a construction similar to that in Tao's proof
of the discrete version~\cite{tao:10}, with the steps
that employed functional submodularity replaced by different
intermediate inequalities that follow from the data
processing property of mutual information. Although both 
these proofs are quite
involved, we note that the submodularity-for-sums
inequality can be used to provide a very short proof
that works in both the discrete and continuous case.

\medskip

\noindent
{\sc Proof of the sum-difference inequality. }
We prove the result for the continuous case; the discrete case 
follows from an identical argument.
Without loss of generality, assume $X$ and $Y$ are independent, 
and let $X'$ be an independent copy of $X$.
Then
    \begin{align*}
        d(X,-Y)
        &= h(X+Y)-\frac{1}{2}h(X)-\frac{1}{2}h(Y)
        \\ &\leq h(X+Y-X')-\frac{1}{2}h(X)-\frac{1}{2}h(Y)
        \\ &\leq h(X-X')+h(Y-X')-\frac{3}{2}h(X)-\frac{1}{2}h(Y)
        \\ &= d(X,X)+d(X,Y)
        \\ &\leq 3d(X,Y).
    \end{align*}
    The three inequalities follow from the fact that convolution
increases differential entropy,
the submodularity for sums bound, 
and the Ruzsa triangle inequality, respectively.
\qed

In the same work~\cite{KM:14}, the following version of the 
Pl\"unnecke--Ruzsa inequality for differential entropy
is established:
Suppose $X,Y_1,Y_2,\dots,Y_n$ are independent 
continuous random variables. If there are constants
$K_1,K_2,\dots,K_n\geq1$ satisfying $h(X+Y_i)\leq h(X)+\log K_i$ for each $i$,
then
\bqq
 h(X+Y_1+Y_2+\cdots+Y_n) \leq h(X)+\log(K_1K_2\cdots K_n). 
\label{eq:PRadd}
\eqq
Using the doubling-difference inequality,
this implies the differential entropy 
analog of~(\ref{eq:HPR}) with 
the explicit factor
$n+2m$ in place of the $O(n+m)$ term.
Once again, we note that the proof immediately 
translates to the discrete case, and the corresponding
inequality holds for discrete random variables.

\section{Additive energy and the differential entropy of sums}
\label{section:entropic_additive_energy}

In this section we define the additive energy $a(X,Y)$ for
continuous random variables $X,Y$, and we establish various
statements that elaborate on the intuition that large
additive energy $a(X,Y)$ corresponds to small entropy
$h(X+Y)$ of the sum of $X$ and $Y$. These results generalize
and strengthen the corresponding bounds for the entropy
of discrete random variables obtained by Goh~\cite{goh:24}.
As the proofs of the discrete bounds in~\cite{goh:24}
rely 
heavily on the functional submodularity property of discrete
entropy, we take a different approach to proving the
corresponding results for differential entropy.

The gist of our approach is the observation that
the combinatorial versions of all the results we consider
in this section follow from the basic 
bound~(\ref{eq:additive_combinatorics_energy_inequality})
described in the Introduction.
Therefore, we begin by establishing its natural
entropic analog in
Lemma~\ref{lemma:entropy_energy_inequality}.
From this, we then derive the continuous
analogs of all the relevant results obtained in~\cite{goh:24}.
This approach has some important advantages,
that indeed apply to almost all the results 
obtained in this paper:

\begin{itemize}
\item
The resulting proofs do not rely on functional
submodularity, so they apply to both the continuous
and discrete cases.
\item
The bounds obtained are often stronger than their
earlier discrete counterpart,
and the slack in the inequalities
can sometimes be precisely quantified.
\end{itemize}
In particular, although our results in this section
are stated for continuous random variables, they all
apply verbatim to discrete random variables as well.

\subsection{Additive energy for continuous random variables}

Recall that, for an arbitrary pair of discrete random variables
$(X,Y)$ with values in an additive group $G$,
we defined 
the entropic additive energy $A(X,Y)$
in~(\ref{eq:AXY}) 
as $A(X,Y)=H(X_1,Y_1,X_2,Y_2,X+Y),$
where 
$(X_1,Y_1)$ and $(X_2,Y_2)$ are conditionally
independent versions of $(X,Y)$, given $X+Y$.
But $Y_1$ and $Y_2$ can be determined from $(X_1,X_2,X+Y)$
so, if $X,Y$ are continuous random variables,
then $h(X_1,Y_1,X_2,Y_2,X+Y)=-\infty$.
On the other hand, the same argument shows that
$A(X,Y)$ is also always equal to $H(X_1,X_2,X+Y)$.
Therefore, we define
the \emph{differential entropic additive energy} of two 
jointly distributed continuous random variables $X$ and $Y$ 
as
$$a(X,Y)=h(X_1,X_2,X+Y),$$
{where
$(X_1,Y_1)$ and $(X_2,Y_2)$ are conditionally
independent versions of $(X,Y)$, given $X+Y$.}

As with the expression for $A(X,Y)$ derived
in~(\ref{eq:AXY2}) in the discrete case, 
we will find the following
alternative representation $a(X,Y)$ useful
and sometimes simpler to work with.

\begin{lemma}
\label{lemma:additive_energy_conditional_formulation}
For any pair $(X,Y)$ of continuous random variables:
    \begin{equation*}
        a(X,Y)=2h(X,Y)-h(X+Y).
    \end{equation*}
\end{lemma}

\noindent
{\sc Proof. }
Let $(X_1,Y_1)$ and $(X_2,Y_2)$ be conditionally 
independent versions of $(X,Y)$ given $X+Y$.
By the chain rule and conditional independence, we have,
    \begin{align*}
        h(X_1,X_2,X+Y)
        &= h(X_1|X+Y)+h(X_2|X+Y)+h(X+Y)
        \\ &= h(X_1,X+Y)-h(X+Y)+h(X_2,X+Y)-h(X+Y)+h(X+Y)
        \\ &= h(X_1,Y_1)+h(X_2,Y_2)-h(X+Y)
        \\ &= 2h(X,Y)-h(X+Y),
    \end{align*}
as claimed.
\qed

As mentioned above, the starting point for all
results in this section will be the following
entropic analog of~\eqref{eq:additive_combinatorics_energy_inequality}.
Recall that the mutual information $I(X;Y)$ between two continuous
random variables $X$ and $Y$ {satisfies}
$I(X;Y)=h(X)+h(Y)-h(X,Y)$,
and similarly in the discrete case.

\begin{lemma}
\label{lemma:entropy_energy_inequality}
    Let $X$ and $Y$ be continuous random variables.
    Then
    \begin{equation}\label{eq:entropy_energy_inequality}
        h(X+Y)+a(X,Y) = 2h(X,Y) \leq 2h(X) + 2h(Y),
    \end{equation}
where the slackness in the inequality is exactly $2I(X;Y)\geq0$.
\end{lemma}

Note that the inequality in~(\ref{eq:entropy_energy_inequality}) appears
to be in the reverse direction compared to its combinatorial
counterpart~(\ref{eq:additive_combinatorics_energy_inequality}).
But when $X$ and $Y$ are independent -- as dictated
by the entropy/cardinality correspondence --
we actually have equality
in~(\ref{eq:entropy_energy_inequality}), which is of course
consistent with~(\ref{eq:additive_combinatorics_energy_inequality}).

\subsection{Large entropic additive energy}

In order to quantify what ``large'' means for additive energy,
note that
$$a(X,Y) = 2h(X,Y) - h(X+Y)
   \leq 2h(X,Y) - h(X|Y)
    = h(X,Y) + h(Y),
$$
so, by symmetry,
\begin{align}\label{eq:additive_energy_upper_bound}
    a(X,Y) \leq h(X,Y) + \min\{h(X),h(Y)\} \leq h(X) 
	+ h(Y) + \min\{h(X),h(Y)\}.
\end{align}
Therefore, we can consider $a(X,Y)$ to be large if
\[ a(X,Y)\geq \frac{3}{2}h(X)+\frac{3}{2}h(Y)-\log C. \]
Similarly, since
$$h(X+Y)
\geq\max\{h(X+Y|Y),h(X+Y|X)\}
=\max\{h(X|Y),h(Y|X)\}
\geq\frac{1}{2}h(X|Y)+\frac{1}{2}h(Y|X),$$ 
we can consider $h(X+Y)$ to be small if
\[ h(X+Y)\leq\frac{1}{2}h(X|Y)+\frac{1}{2}h(Y|X)+\log C. \]
The following two results show that large $a(X,Y)$ implies small $h(X+Y)$, 
with a partial converse if $X$ and $Y$ are only weakly dependent.
{Corollary~\ref{cor:large_energy_small_sum} is
the differential entropy version of 
the discrete entropy result in \cite[Proposition~3]{goh:24}.}


\begin{corollary}
\label{cor:large_energy_small_sum}
    Let $X,Y$ be continuous random variables.
    If
    \begin{equation}\label{eq:large_energy_small_sum_1}
        a(X,Y) \geq \frac{3}{2}h(X)+\frac{3}{2}h(Y)-\log C,
    \end{equation}
    for some constant $C$, then
    \begin{equation}\label{eq:large_energy_small_sum_2}
        h(X+Y) \leq \frac{1}{2}h(X)+\frac{1}{2}h(Y)+\log C.
    \end{equation}
	Conversely, if~\eqref{eq:large_energy_small_sum_2} holds
	and $X$ and $Y$ are weakly dependent in the sense that
    \begin{equation}\label{eq:large_energy_small_sum_3}
        h(X,Y)\geq h(X)+h(Y)-\log C',
    \end{equation}
	for some constant $C'$, then 
    \begin{equation}\label{eq:large_energy_small_sum_4}
        a(X,Y) \geq \frac{3}{2}h(X)+\frac{3}{2}h(Y)-\log C - 2\log C'.
    \end{equation}
    In particular, if $X$ and $Y$ are independent, then~\eqref{eq:large_energy_small_sum_2} implies~\eqref{eq:large_energy_small_sum_1}.
\end{corollary}

\noindent
{\sc Proof. }
By Lemma~\ref{lemma:entropy_energy_inequality},
the bound~\eqref{eq:large_energy_small_sum_1} 
implies~\eqref{eq:large_energy_small_sum_2}.
Conversely, by
~\eqref{eq:large_energy_small_sum_2},~\eqref{eq:large_energy_small_sum_3},
and Lemma~\ref{lemma:entropy_energy_inequality},
we have
    \begin{align*}
        2h(X) + 2h(Y)-2\log C' 
        &\leq 2h(X,Y) 
        \\ &= a(X,Y) + h(X+Y)
        \\ &\leq a(X,Y) + \frac{1}{2}h(X)+\frac{1}{2}h(Y)+\log C,
    \end{align*}
    and rearranging yields~\eqref{eq:large_energy_small_sum_4}.
\qed

{ 
If, instead of using Lemma~\ref{lemma:entropy_energy_inequality},
we use the mutual information $I(X;Y)$,
we immediately obtain the following 
strengthening 
of Corollary~\ref{cor:large_energy_small_sum}
and of the corresponding discrete
entropy result in \cite[Proposition~3]{goh:24}.}

\begin{corollary}
\label{cor:improved_large_energy_small_sum}
    Let $X$ and $Y$ be continuous random variables and $C$ be a constant.
    Then
    \begin{equation*}
        a(X,Y) \geq \frac{3}{2}h(X) + \frac{3}{2}h(Y) - \log C,
    \end{equation*}
    if and only if,
    \begin{equation*}
        h(X+Y) \leq \frac{1}{2}h(X) + \frac{1}{2}h(Y) + \log C - 2I(X;Y).
    \end{equation*}
\end{corollary}

\noindent
{{\sc Proof. }
From the definitions we have 
\begin{align*}
a(X,Y)-\frac{3}{2}h(X)-\frac{3}{2}h(Y)
&=
	2h(X,Y)-h(X+Y)
	-\frac{3}{2}h(X)-\frac{3}{2}h(Y)\\
&=
	\frac{1}{2}h(X)+\frac{1}{2}h(Y)-2[h(X)+h(Y)-h(X,Y)]-h(X+Y)\\
&=
	\frac{1}{2}h(X)+\frac{1}{2}h(Y)-2I(X;Y)-h(X+Y),
\end{align*}
which gives the claimed equivalence.
\qed}

Next, we show that either of the statements in the above two corollaries --
namely, $a(X,Y)$ being large in the 
sense of~(\ref{eq:large_energy_small_sum_1}),
or $h(X+Y)$ being small 
in the sense of~(\ref{eq:large_energy_small_sum_2})
--
is equivalent to the entropies $h(X)$ and $h(Y)$ being close in value.
The following result generalizes and strengthens
\cite[Proposition~7]{goh:24} for discrete entropy.

\begin{corollary}
\label{cor:symmetric_entropy_control}
    Let $X$ and $Y$ be continuous random variables and $C$ be a constant.
    Then we have
    \begin{equation}\label{eq:symmetric_entropy_control_1}
        a(X,Y) \geq \frac{3}{2}h(X) + \frac{3}{2}h(Y) - \log C
    \end{equation}
	if and only if
    \begin{equation}\label{eq:symmetric_entropy_control_2}
        h(X) \leq h(Y) + 2\log C - 2I(X+Y;Y) + 2I(X;Y),
    \end{equation}
	or, equivalently,
    \begin{equation}\label{eq:symmetric_entropy_control_3}
        h(Y) \leq h(X) + 2\log C - 2I(X+Y;X) + 2I(X;Y).
    \end{equation}
    Under any, hence all, of these conditions,
    \begin{equation}\label{eq:symmetric_entropy_control_4}
        h(Y) - 2\log C + 2I(X+Y;X) - 2I(X;Y) \leq h(X) 
	\leq h(Y) + 2\log C - 2I(X+Y;Y) + 2I(X;Y).
    \end{equation}
In particular, 
    \begin{equation}\label{eq:symmetric_entropy_control_5}
        \abs{h(X)-h(Y)} \leq 2\log C + 2I(X;Y) - 2\min\{I(X+Y;X),I(X+Y;Y)\}.
    \end{equation}
\end{corollary}

\noindent
{\sc Proof. }
Using the bound~\eqref{eq:symmetric_entropy_control_1} 
in the equality part of Lemma~\ref{lemma:entropy_energy_inequality} yields
    \begin{align*}
        \frac{3}{2}h(X) + \frac{3}{2}h(Y) - \log C + h(X+Y) \leq h(X) + h(Y) + h(X|Y) + h(Y|X).
    \end{align*}
Since
$h(X+Y)=h(X|Y)+I(X+Y;Y)$ and $h(Y|X)= h(Y)-I(X;Y)$, we obtain,
    \begin{align*}
        \frac{3}{2}h(X) + \frac{3}{2}h(Y) - \log C + I(X+Y;Y) \leq h(X) + 2h(Y) - I(X;Y),
    \end{align*}
and, rearranging, gives~\eqref{eq:symmetric_entropy_control_2}.
As we only combined the assumed bound~(\ref{eq:symmetric_entropy_control_1}) with 
equalities, reversing the argument immediately 
shows that~\eqref{eq:symmetric_entropy_control_2} 
implies~\eqref{eq:symmetric_entropy_control_1}.
The equivalence of~\eqref{eq:symmetric_entropy_control_1} 
with~\eqref{eq:symmetric_entropy_control_3} follows by symmetry.
    Combining~\eqref{eq:symmetric_entropy_control_2} and~\eqref{eq:symmetric_entropy_control_3} yields~\eqref{eq:symmetric_entropy_control_4}, which 
implies~\eqref{eq:symmetric_entropy_control_5}.
\qed

The reason why large additive energy $a(X,Y)$ implies 
$h(X)$ and $h(Y)$ are close is, in part, because
our definition of what it means for $a(X,Y)$ to be
large is symmetric in $X$ and $Y$.

In the asymmetric case where $h(X)$ and $h(Y)$ 
are significantly different, with $h(X) < h(Y)$, say,
then in view of~\eqref{eq:additive_energy_upper_bound} 
we can instead consider $a(X,Y)$ to be large 
if $a(X,Y) \geq 2h(X)+h(Y)-\log C$.
Similarly, since $h(X+Y)\geq h(Y|X)$, we can 
consider $h(X+Y)$ to be small if $h(X+Y)\leq h(Y|X)+\log C$.
The following result states that, if $a(X,Y)$ is large in this
sense, then $h(X+Y)$ is appropriately small, with 
a partial converse when $X$ and $Y$ are weakly
dependent. This strengthens and 
generalizes~\cite[Proposition~8]{goh:24}
to the continuous case.
The proof of 
Corollary~\ref{cor:improved_asymmetric_large_energy_small_sum}
is essentially identical to that of 
Corollary~\ref{cor:improved_large_energy_small_sum},
so we omit it.


\begin{corollary}
\label{cor:improved_asymmetric_large_energy_small_sum}
    Let $X$ and $Y$ be continuous random variables and $C$ be a constant.
    Then
    \[ a(X,Y) \geq 2h(X) + h(Y) - \log C, \]
    if and only if
    \[ h(X+Y) \leq h(Y) + \log C - 2I(X;Y)\leq h(Y|X)+\log C. \]
\end{corollary}

Finally, the following result can be viewed as a
version of the contrapositive of 
Corollary~\ref{cor:large_energy_small_sum}.
It states that, if the entropy $h(X+Y)$ is large, then 
the entropic additive energy $a(X,Y)$ is small, 
again with a partial converse when $X$ and $Y$ 
are only weakly dependent.
Corollary~\ref{cor:improved_large_sum_small_energy},
which 
follows immediately from Lemma~\ref{lemma:entropy_energy_inequality},
strengthens~\cite[Proposition~10]{goh:24} and
generalizes it to the continuous case.


\begin{corollary}\label{cor:improved_large_sum_small_energy}
    Let $X$ and $Y$ be continuous random variables and $C$ be a constant.
    Then,
    \[ h(X+Y)\geq h(X)+h(Y)+\log C, \]
    if and only if,
    \[ a(X,Y)\leq h(X) + h(Y) + \log C - 2I(X;Y). \]
\end{corollary}

\section{The BSG theorem for differential entropy}
\label{section:BSG}

Here we establish an entropic version of
an important result in additive
combinatorics, known as the 
Balog--Szemer\'edi--Gowers (BSG) theorem~\cite{balog:94,gowers:98}.
The BSG theorem further
quantifies the intuition described earlier, that 
sets $A$ and $B$ with large additive energy $E(A,B)$ must contain 
sizeable additive structure. Specifically, 
it states that if $E(A,B)$ is large, then
$A$ and $B$ must contain high-density subsets $A'$ and $B'$ 
such that $\abs{A'+B'}$ is small.
It can be viewed as a partial converse 
to the observation~\eqref{old_eq:additive_energy_symmetrized} 
that a small sumset $\abs{A+B}$ implies a large additive energy $E(A,B)$.
Note that a full converse does not hold in general; there exist sets 
$A$ and $B$ with large $E(A,B)$ but small $\abs{A+B}$,
so the conclusion of the BSG theorem
is best possible, up to quantitative constants.

\begin{theorem}[Combinatorial BSG theorem]
    Let $A$ and $B$ be finite subsets of the same additive group, and suppose $E(A,B) \geq c\abs{A}^{3/2}\abs{B}^{3/2}$ for some constant $c$.
    Then there are positive constants $c'$, $c''$, and $C$ depending only on $c$ such that there exist $A'\subset A$ and $B'\subset B$ with $\abs{A'}\geq c'\abs{A}$ and $\abs{B'}\geq c''\abs{B}$ satisfying
    \[ \abs{A'+B'}\leq C\abs{A}^{1/2}\abs{B}^{1/2}. \]
\end{theorem}

An entropic BSG theorem 
was established by Tao in \cite[Theorem~3.1]{tao:10},
and its differential entropy analog was proved 
in~\cite[Theorem~3.14]{KM:14}.
More recently,
Gowers, Green, Manners, and Tao~\cite[Lemma~A.2]{GGMT:23b} 
proved a similar result with better constants
for discrete entropy.
However, the most faithful entropy analog of the 
BSG theorem is due to Goh~\cite{goh:24},
which is also the only one stated in terms of entropic additive energy.

Since the natural probabilistic analog of restricting to a subset 
is conditioning, all entropic BSG theorems establish
the existence of particular conditionings that lead to
appropriate bounds. Specifically, 
suppose the additive energy $A(X,Y)$ is large
in the sense of~(\ref{eq:AXYequivalent}),
and consider conditionally independent versions 
$(X_1,Y_1)$ and $(X_2,Y_2)$. Then
Goh's~\cite[Theorem~6]{goh:24} says that
the conditional entropies
$H(X_1|X+Y)$ and $H(Y_2|X+Y)$
are close to $H(X)$ and $H(Y)$, respectively,
while the conditional entropy of the sum,
$H(X_1+Y_2|X+Y)$, is small, again in 
the sense of~(\ref{eq:AXYequivalent}).

In this section, we state and prove the differential entropy analog 
of Goh's BSG theorem. Although the proof in~\cite{goh:24}
relies on functional submodularity and hence does not 
generalize to the case of differential entropy, our proof
of Theorem~\ref{theorem:bsg} below applies verbatim to the
discrete case as well. We begin by establishing
a useful upper bound on the additive energy. The discrete
version of Lemma~\ref{lemma:bsg_lemma}
first appeared in~\cite{GGMT:23b}.

\begin{lemma}
\label{lemma:bsg_lemma}
    Suppose $X$ and $Y$ are continuous random variables,
and let $(X_1,Y_1)$ and $(X_2,Y_2)$ be conditionally independent 
versions of $(X,Y)$ given $X+Y$.
    Then
    \begin{equation}\label{eq:bsg_lemma}
        \max\{h(X_1-X_2),h(X_1-Y_2)\} \leq 2h(X)+2h(Y)-a(X,Y).
    \end{equation}
\end{lemma}

\noindent
{\sc Proof. }
We first prove the inequality for the $h(X_1-Y_2)$ term.
{ 
By our assumptions,
$X_1+Y_1=X_2+Y_2$, so,
\begin{align*}
X_1-Y_2
&=
	(X_1+Y_1)-Y_1-Y_2\\
&=
	(X_2+Y_2)-Y_1-Y_2\\
&=
	X_2-Y_1.
\end{align*}
} And since for arbitrary random variables $A,B$ and $C$ we
always have, $h(A|B,C)\leq h(A|B)$, we can bound
    \begin{align*}
        h(X_1,Y_1,X_1-Y_2) + h(X_1-Y_2)
        &= h(X_1,X_1-Y_2) + h(Y_1|X_1,X_1-Y_2)+h(X_1-Y_2)
        \\ &\leq h(X_1,X_1-Y_2) + h(Y_1|X_1-Y_2)+h(X_1-Y_2)
        \\ &= h(X_1,X_1-Y_2) + h(Y_1,X_1-Y_2)
        \\ &= h(X_1,Y_2)+h(Y_1,X_2-Y_1)
        \\ &\leq h(X)+h(Y)+h(Y_1,X_2)
        \\ &\leq 2h(X)+2h(Y).
    \end{align*}
    Thus,
    \begin{align*}
        h(X_1-Y_2)
        &\leq 2h(X)+2h(Y)-h(X_1,Y_1,X_1-Y_2),
    \end{align*}
    so it suffices to show
    \[ h(X_1,Y_1,X_1-Y_2)=a(X,Y). \]
    But this immediately follows from noting that 
    \[ h(X_1,Y_1,X_1-Y_2)=h(X_1,Y_1,Y_2)=h(X_1,X_2,X+Y)=a(X,Y). \]
    
Similarly, for the $h(X_1-X_2)$ term, we have 
    \begin{align*}
        h(X_1,Y_1,X_1-X_2)+h(X_1-X_2)
        &= h(X_1,X_1-X_2)+h(Y_1|X_1,X_1-X_2)+h(X_1-X_2)
        \\ &\leq h(X_1,X_2)+h(Y_1|X_1-X_2)+h(X_1-X_2)
        \\ &= h(X_1,X_2)+h(Y_1,X_1-X_2)
        \\ &= h(X_1,X_2)+h(Y_1,Y_2-Y_1)
        \\ &= h(X_1,X_2)+h(Y_1,Y_2)
        \\ &\leq 2h(X)+2h(Y).
    \end{align*}
    Thus it again suffices to 
show $h(X_1,Y_1,X_1-X_2)=a(X,Y)$, which follows from observing
$$ h(X_1,Y_1,X_1-X_2)=h(X_1,Y_1,X_2)=h(X_1,X_2,X+Y)=a(X,Y). $$
Combining the two bounds yields~(\ref{eq:bsg_lemma}).
\qed

{
As alluded to in~\cite{GGMT:23b}, 
Lemma~\ref{lemma:bsg_lemma} allows one 
to prove a better version of the entropic BSG theorem;
the original entropic BSG was proved
in the discrete case by Tao
in~\cite[Theorem~3.1]{tao:10}
and in the continuous case by
Kontoyiannis and Madiman 
in~\cite[Theorem~3.14]{KM:14}.}

\begin{theorem}[Entropic BSG theorem]
\label{theorem:bsg}
Suppose the continuous random variables $X$ and $Y$ satisfy,
    for some constant $C$,
    \[ a(X,Y) \geq \frac{3}{2}h(X)+\frac{3}{2}h(Y)-\log C.\]
Then, taking $(X_1,Y_1)$ and $(X_2,Y_2)$ to be conditionally independent 
versions of $(X,Y)$ given $X+Y$, we have
    \begin{align}
h(X_1|X+Y) 
&\geq 
	h(X) - 2\log C,
	\label{eq:entropic_BSG_1}\\
h(Y_2|X+Y)
&\geq 
	h(Y) - 2\log C,
	\label{eq:entropic_BSG_2}
\end{align}
and, moreover,
$X_1$ and $Y_2$ are conditionally independent given $X+Y$, 
and satisfy
    \begin{equation}\label{eq:entropic_BSG_3}
        h(X_1+Y_2|X+Y) \leq \frac{1}{2}h(X)+\frac{1}{2}h(Y)+\log C.
    \end{equation}
\end{theorem}

\noindent
{\sc Proof. }
    With $(X,Y),(X_1,Y_1),(X_2,Y_2)$ as in the theorem statement, 
	{and using 
	Lemma~\ref{lemma:additive_energy_conditional_formulation}},
	we have
    \begin{align*}
        h(X_1|X+Y)
        &= h(X_1,X+Y) - h(X+Y)
        \\ &= h(X_1,Y_1)-h(X+Y)
        \\ &= a(X,Y)-h(X,Y)
        \\ &\geq \frac{3}{2}h(X)+\frac{3}{2}h(Y)-\log C - h(X,Y)
        \\ &\geq \frac{1}{2}h(X)+\frac{1}{2}h(Y)-\log C.
    \end{align*}
    We similarly obtain
    \[ h(Y_2|X+Y) \geq \frac{1}{2}h(X)+\frac{1}{2}h(Y)-\log C. \]
    Summing these two bounds yields
    \begin{align*}
        h(X_1|X+Y)+h(Y_2|X+Y) \geq h(X)+h(Y)-2\log C,
    \end{align*}
    from which it follows that
    \begin{align*}
        h(X_1|X+Y) 
        & \geq h(X)+h(Y)-h(Y_2|X+Y)-2\log C
        \\ &\geq h(X)-2\log C.
    \end{align*}
	This proves~(\ref{eq:entropic_BSG_1}),
	and~(\ref{eq:entropic_BSG_2}) follows by symmetry.

Since $(X_1+Y_2)-(X+Y)=X_1-X_2$, we have
    \begin{align*}
        h(X_1+Y_2|X+Y)
        &= h(X_1+Y_2,X+Y) - h(X+Y)
        \\ &= h(X_1-X_2,X+Y) - h(X+Y)
        \\ &= h(X_1-X_2|X+Y)
        \\ &\leq h(X_1-X_2),
    \end{align*}
and using Lemma~\ref{lemma:bsg_lemma}, 
    \begin{align*}
        h(X_1+Y_2|X+Y)
        &\leq 2h(X) + 2h(Y)-a(X,Y)
        \\ &\leq \frac{1}{2}h(X)+\frac{1}{2}h(Y)+\log C.
    \end{align*}
This gives~\eqref{eq:entropic_BSG_3} and completes the proof.
\qed

As observed by Goh~\cite{goh:24},
{the assumptions of 
the BSG theorems in
\cite[Theorem~3.1]{tao:10}
and \cite[Theorem~3.14]{KM:14}}
combined with
Corollary~\ref{cor:large_energy_small_sum} 
imply that
\[ a(X,Y) \geq \frac{3}{2}h(X)+\frac{3}{2}h(Y)-3\log K, \]
and Theorem~\ref{theorem:bsg} then 
gives~\eqref{eq:entropic_BSG_1}--\eqref{eq:entropic_BSG_3}
with error terms $-6\log K$, $-6\log K$, and $3\log K$, respectively.
Meanwhile, the analogous statements in
the BSG statements of~\cite{tao:10} and~\cite{KM:14}
have error 
terms $-\log K$, $-\log K$, and $7\log K$, respectively.
Thus, Theorem~\ref{theorem:bsg} has worse bounds on the conditional 
entropies of $X_1$ and $Y_2$, but better bounds on 
the ``main'' inequality~\eqref{eq:entropic_BSG_3} on the 
conditional entropy of the conditionally independent sum $X_1+Y_2$.

\section{Stability of large discrete entropic doubling}
\label{section:Sidon_stability}

Before continuing with our development
of differential entropy inequalities,
in this section we temporarily return 
to the case of discrete entropy.
For a discrete random variable $X$ taking values in a
subset $A$ of an additive group $G$, we define
the {\em doubling constant} $s(X)$ in analogy 
with the combinatorial doubling constant $s(A)$ 
in~(\ref{eq:doublingA}), via,
\bqq
s(X)=H(X+X')-H(X),
\label{eq:doublingH}
\eqq
where $X,X'$ are i.i.d.\
Clearly, $s(X)\geq 0$, and Tao~\cite{tao:10}
showed that $s(X)$ is small if and only 
if $X$ is approximately uniformly distributed
on a generalized arithmetic progression.
Here, we examine the behaviour of the
doubling constant $s(X)$ at its other extreme.
We note that $s(X)$ is always bounded above by
$H(X)$, and we show that $s(X)$ is close to $H(X)$
if and only if $X$ is approximately supported on
a {\em Sidon set}.
Recall that $A$ is a \emph{Sidon set} if, 
whenever $a+b=c+d$ 
for $a,b,c,d\in A$, we necessarily
have $\set{a,b}=\set{c,d}$.

Let $X$ be a discrete random variable with
probability mass function $P$ on $A$.
Goh~\cite{goh:24} observed that,
if $X$ is supported on a Sidon set,
i.e., if $\{a\in G:P(a)>0\}$
is a Sidon set, then
$s(X)\geq H(X)-\log 2$.
Our first result identifies a tighter upper bound
to $s(X)$ {than the trivial bound $H(X)$},
and shows that it is achieved if
and only if 
$X$ is supported on a Sidon set.

\begin{lemma}
\label{lemma:Sidon_bound}
If $X$ is discrete random variable with values 
in a subset $A$ of an additive group $G$,
and with probability mass function $P$,
then,
    \begin{equation}
       s(X) \leq H(X)-(\log 2) \left(1-\sum_{a\in A} P(a)^2\right),
	\label{eq:Sidon_bound}
    \end{equation}
    with equality if and only if
$X$ is supported on 
a Sidon set.
\end{lemma}

\noindent
{\sc Proof. }
By the definition of $s(X)$,
\begin{align*}
    H(X) - s(X)
    &= 2H(X)-H(X+X')
    \\ &= H(X,X'|X+X')
    \\ &= \sum_{a,b\in A} P(a)P(b)\log
	\Big(\frac{\sum_{ c\in A} P(c)P(a+b-c)}{P(a)P(b)}\Big).
\end{align*}
This expression together with the observation that
\bqq
\sum_{c\in A} P(c) P(a+b-c) \geq P(a)P(b)+P(b)P(a) = 2P(a)P(b),
\label{eq:sidon2}
\eqq
whenever $a\neq b$, immediately imply~(\ref{eq:Sidon_bound}).

Conversely, in order to have
equality in~(\ref{eq:Sidon_bound})
we must have equality in~(\ref{eq:sidon2}),
which can only happen 
if any possible value $a+b$ of the sum $X+X'$ that has
nonzero probability can only be achieved by $(X,X')=(a,b)$
or $(X',X)=(a,b)$; this implies that $X$ is supported on
a Sidon set.
\qed

We now prove a stability result, showing that if $s(X)$
is close to its upper bound in~\eqref{eq:Sidon_bound}, 
then $X$ is close to being supported on a Sidon set.

\begin{proposition}
\label{proposition:Sidon_stability}
    Let $X$ be a discrete random variable supported
	on a finite subset $A$ of an additive group $G$, 
	and with probability mass function $P$.
    If
    \bqq
	s(X) \geq H(X)- (\log 2)\left(1-\sum_{a\in A} P(a)^2\right)-C,
	\label{eq:Sidoncond}
    \eqq
    for some constant $C\geq0$, then there exists a 
Sidon set $B\subset A$ such that
	\bqq
     \BBP(X\in B) \geq 1-\frac{C}{p_*(\log 2)},
	\label{eq:inSidon}
	\eqq
where $p_*=\min\{P(a):a\in A\}>0$.
\end{proposition}

\noindent
{\sc Proof. }
    For $a,b\in A$, let
    \[ R(a,b) = \log\Big(\frac{\sum_{c\in A} P(c)P(a+b-c)}{P(a)P(b)}\Big)
 - (\log 2)\IND_{\{a\neq b\}} \geq 0. \]
{To see that $R(a,b)$ is nonnegative}
note that, for any $a,b\in A$,
$$\sum_{c\in A}P(c)P(a+b-c)\geq \frac{2P(a)P(b)}{1+\IND_{\{a=b\}}}.$$
    Then our assumption is equivalent to the inequality 
$\BBE[R(X,X')]\leq C$, where $X$ and $X'$ are i.i.d.,
    and by Markov's inequality, we have
    \begin{equation}\label{eq:Sidon_Markov_bound}
        \BBP(R(X,X')\geq\log 2) \leq \frac{C}{\log 2}.
    \end{equation}

    We construct a subset $D\subset A$ as follows.
    Let $I$ be the set of unordered pairs $\{a,b\}$ 
	with $R(a,b)\geq\log 2$, noting that $R(a,b)=R(b,a)$ 
	so this is a well-defined set.
    For any $\{a,b\}\in I$,
	if $P(a)\neq P(b)$, then include in $D$ 
	the element with the smaller probability;
	otherwise, if $P(a)=P(b)$, include in $D$
	one of the two, choosing arbitrarily.
    Let $B=A\setminus D$.

    We first prove~(\ref{eq:inSidon}).
    By~\eqref{eq:Sidon_Markov_bound} and the definition of $D$, we have
    \begin{align*}
        \BBP(X\in D) 
        &\leq \sum_{\{a,b\}\in I}\min\{P(a),P(b)\}
        \\ &\leq \sum_{\{a,b\}\in I}\frac{P(a)P(b)}{p_*}
	\left(1+\IND_{\{a\neq b\}}\right)
        \\ &= \frac{1}{p_*}\BBP(R(X,X')\geq\log 2)
        \\ &\leq \frac{C}{p_*(\log 2)},
    \end{align*}
which is equivalent to~\eqref{eq:inSidon}.

It remains to show $B$ is a Sidon set.
    Notice that for all $a,b\in B$, we have $R(a,b)<\log 2$.
    Suppose $B$ is not a Sidon set, i.e., that there are
	$a_1,b_1,a_2,b_2\in B$ are such that 
    $a_1+b_1=a_2+b_2$ and $\{a_1,b_1\}\neq\{a_2,b_2\}$. 
    This implies that,
	\bqq
	\sum_{c\in A}P(c)P(a_1+b_1-c)
	\geq 
	\frac{2P(a_1)P(b_1)}{1+\IND_{\{a_1=b_1\}}}
	+\frac{2P(a_2)P(b_2)}{1+\IND_{\{a_2=b_2\}}}.
	\label{eq:preSidon}
	\eqq
    Now consider two cases.
	If
    \[ \frac{P(a_1)P(b_1)}{1+\IND_{\{a_1=b_1\}}} 
	\leq \frac{P(a_2)P(b_2)}{1+\IND_{\{a_2=b_2\}}}, \]
	then~(\ref{eq:preSidon}) and the definition of $R(a,b)$
	imply that $R(a_1,b_1)\geq\log 2$,
	while if the reverse inequality holds,
	 we have $R(a_2,b_2)\geq\log 2$,
	a contradiction either way.
      	Hence, $B$ is a Sidon set, and the proof is complete.
\qed

The following two examples 
demonstrate that
the assumption that $X$ takes on only finitely 
many values -- equivalently, that $p_*$ is strictly positive -- 
cannot be removed. In fact, the dependence of the 
bound~\eqref{eq:inSidon} on $p_*$ cannot
be entirely avoided. 

\begin{example}
Consider an $N$-element
set $A\subset G$ that has exactly one violation 
to the Sidon set condition, that is,
exactly one nontrivial solution to $a_1+b_1=a_2+b_2$,
and suppose also that, in that case,
both $a_1\neq b_1$ and $a_2\neq b_2$.
Such an $A$ can be easily constructed in $G=\IN$, for example.
Let $X$ be the uniform distribution on $A$.
Then we see that
    \[ s(X) \geq 
	H(X) -(\log 2)\left(1-\sum_{a\in A} P(a)^2\right)
	-\frac{4\log 2}{N^2},
    \]
so the assumption of Proposition~\ref{proposition:Sidon_stability}
holds with $C=\frac{4\log 2}{N^2}$, 
    while for any Sidon set $B\subset G$ we have
    \[ \BBP(X\in B) \leq \frac{N-1}{N} = 1 - \frac{1}{N}
	=1-\frac{C}{4p_*(\log 2)}. \]
\end{example}

\begin{example}
    A more striking example is as follows.
    Consider the $4N$-element subset $A\subset\IN$ given by
    \[ A = \bigcup_{k=0}^{N-1} 
	\big\{10^k,2\cdot 10^k,4\cdot 10^k,5\cdot10^k\big\}. \]
    It is easy to see that the only violations to the Sidon set condition are of the form
    \[ 10^k + 5\cdot 10^k = 2\cdot 10^k + 4\cdot 10^k. \]
    Thus, letting $X$ be the uniform distribution on $A$, we have
    	\[ s(X) \geq H(X)
	-(\log 2)\left(1-\sum_{a\in A} P(a)^2\right) 
	-\frac{\log 2}{4N},
	\]
    yet any Sidon set $B\subset \IN$ can contain at 
	most 3 of the 4 elements 
	of $\{10^k,2\cdot10^k,4\cdot10^k,5\cdot10^k\}$, which implies
    \[ \BBP(X\in B) \leq \frac{3}{4}=1-\frac{1}{4}. \]
    In other words, 
	as $N\to\infty$, 
	the condition of 
	Proposition~\ref{proposition:Sidon_stability}
	holds with
	$C=\frac{\log 2}{4N}\to0$,
	while for any Sidon set $B$ the probability
	$\BBP(X\in B)$ stays bounded away from~1.
	Still, this does not violate the result
	of Proposition~\ref{proposition:Sidon_stability},
	because here we have $\frac{C}{p_*(\log 2)}=1$.
\end{example}

Proposition~\ref{proposition:Sidon_stability} and the
two examples above raise the following question.

\medskip

\noindent
{\bf Open problem. } It would be interesting to determine
whether there exists a function $f(C,D)\geq0$ for $C,D\geq0$, 
with $f(C,D)\to0$ as $C\to0$ for any fixed $D\geq0$, such that 
	the following holds: 
If $X$ is a discrete random variable 
	{with values in $A$ and}
	with $H(X)\leq D$ and satisfying 
	condition~\eqref{eq:Sidoncond}
	of Proposition~\ref{proposition:Sidon_stability}
	with constant $C$,
	then there exists a Sidon set $B\subset A$ such that
    	$\BBP(X\in B) \geq 1-f(C,D).$

\medskip

Finally, we return to the obvious upper bound
$s(X)\leq H(X)$.  Lemma~\ref{lemma:Sidon_bound} 
shows that this is achieved if and only if 
$X$ is deterministic. Our last result
in this section is a stability statement,
showing that if $s(X)$ is close to $H(X)$, 
then $X$ is close to being deterministic.

\begin{proposition}
    Let $X$ be a discrete random variable with values
in a subset $A$ of an additive group $G$,
and with probability mass function $P$.
    If
    \[ s(X) \geq H(X)-\varepsilon \]
    for some constant $\varepsilon>0$, then
    \[ p^*:=\max_{a\in A}P(a) \geq 1 - \frac{\varepsilon}{\log 2}. \]
\end{proposition}

\noindent
{\sc Proof. }
    As
    \[ \sum_{a\in A} P(a)^2 \leq p^* \sum_{a\in A}P(a) = p^*, \]
    by Lemma~\ref{lemma:Sidon_bound} we have
    \[ 1-p^* \leq 1 - \sum_{a\in A}P(a)^2 \leq \frac{H(X)-s(X)}{\log 2} 
	\leq \frac{\varepsilon}{\log 2}, \]
    which gives the claimed inequality.
\qed

\section{Differential entropy bounds for products}
\label{section:product}

In this section 
we describe how all the {\em additive} entropy inequalities
for sums of continuous random variables
discussed in Section~\ref{s:diffh} have natural
and easy to derive {\em multiplicative} analogs.

\subsection{Multiplicative entropy}

Let $X$ be an arbitrary continuous random variable.
Since $\BBP(X=0)=0$,
we can think of $X$ as
taking values in the group $\RL^\times:=\RL\setminus\{0\}$
equipped with the multiplication operation.
Recall that
$(\RL^\times,\cdot)$ is an abelian 
group with Haar measure $\rho$ given by
its density $\frac{d\rho}{d\lambda}(x)=\frac{1}{|x|}$,
$x\in\RL^\times$, with respect to Lebesgue measure $\lambda$.
If
$X$ has law $\mu$ with density 
$f=d\mu/d\lambda$ with respect to Lebesgue measure,
then it also has density 
$$g(x)=\frac{d\mu}{d\rho}(x)=|x|f(x),\quad x\in\RL^\times,$$
with respect to Haar measure.

In~\cite{KM:16} it was observed that a number of
entropy inequalities for sums,
including those mentioned in Section~\ref{s:diffh}, 
also hold for random variables taking values
in arbitrary locally compact, Polish, abelian groups $(G,*)$,
with addition replaced by the group operation $*$,
and with differential entropy replaced by its obvious
analog,
$$\widetilde{h}(X)=-\int g\log g\,d\rho,$$
where $\rho$ denotes the Haar measure on 
the Borel $\sigma$-algebra of $G$.
In the case of $(\RL^\times,\cdot)$,
we call $\widetilde{h}(X)$ the {\em multiplicative
entropy} of a continuous random variable $X$,
and observe that, if $X$ has density $f$ with
respect to Lebesgue measure, then
$\widetilde{h}(X)$ and the usual differential
entropy $h(X)$ are related via
\bqq
 \widetilde h(X) = -\int g\log g\, d\rho = -\int \abs{x} f(x) 
\log(\abs{x}f(x)) \,\frac{dx}{\abs{x}} = h(X) - \BBE[\log\abs{X}]. 
\label{eq:entent}
\eqq
The joint multiplicative entropy of a finite collection of continuous
random variables is defined in the obvious way, and we similarly define
the conditional multiplicative entropy, the multiplicative
mutual information, and so on.

Here and in Section~\ref{s:multisX} we note a number
of ``multiplicative'' entropy inequalities that we will
find useful in Section~\ref{section:sum_product}.
All these results 
follow from the correspondence noted in~\cite{KM:16}
and, except for the simple computations
in Lemmas~\ref{lemma:differential_entropy_inverse}
and~\ref{lemma:differential_entropy_random_multiplication},
they are stated without proof.
In all of our results, we implicitly 
assume that any differential entropies
and multiplicative entropies
appearing in the statement exist and are finite.
Nevertheless, for the sake of clarity, we 
usually explicitly state the integrability
assumptions required for the multiplicative
entropy to be finite.

Just as differential entropy is translation invariant,
$h(X+c)=h(X)$, multiplicative entropy is scale invariant:

\begin{lemma}
\label{lemma:differential_entropy_inverse}
For any continuous random variable $X$ such that $\log |X|$ is integrable,
    \[ h(1/X) = h(X) - 2\BBE[\log\abs{X}], \]
    or, equivalently,
    \[ \widetilde h(1/X) = \widetilde h(X). \]
\end{lemma}

\noindent
{\sc Proof. }
Let $f_X,f_{1/X}$ denote the densities of $X$ and $1/X$ 
with respect to Lebesgue measure.
{Since $f_{1/X}(y)=(1/y^2)f_X(y)$, $y\neq 0$,} we have
    \begin{align*}
        h(1/X)
        &= -\int_\RL f_{1/X}(y)\log f_{1/X}(y) \,dy
        \\ &= -\int_\RL \frac{1}{y^2}f_X(1/y) \log\left(\frac{1}{y^2}f_X(1/y)\right)\,dy
        \\ &= -\int_\RL f_X(x)\log(x^2 f_X(x))\,dx
        \\ &= h(X) - 2\BBE[\log\abs{X}], 
    \end{align*}
{where we note that the reversal of the range of integration
``cancels out'' with the negative sign from the change of variables.}
The second identity follows from~(\ref{eq:entent}).
\qed

\begin{lemma}
\label{lemma:differential_entropy_random_multiplication}
	{
    Let $(X_1,\ldots,X_n,Y)$ be a continuous random vector, 
such that $\log|Y|$
is integrable.
Then, for all integers $k_1,k_2,\dots,k_n$,
    \begin{equation}
	\label{eq:differential_entropy_random_multiplication}
        h(X_1 Y^{k_1},X_2Y^{k_2},\dots,X_nY^{k_n}|Y) = h(X|Y) + 
	K \BBE[\log\abs{Y}],
    \end{equation}
	where $X=(X_1,\ldots,X_n)$} and $K=k_1+\cdots+k_n$. Equivalently,
    \[ \widetilde h(X_1Y^{k_1},X_2Y^{k_2},\dots,X_nY^{k_n}|Y) 
	= \widetilde h(X|Y). \]
\end{lemma}

\noindent
{\sc Proof. }
Let
$Z=(X_1 Y^{k_1},X_2Y^{k_2},\dots,X_nY^{k_n})$.
	Using the obvious notation for joint and conditional 
	densities and writing 
	$z'=(z_1/y^{k_1},z_2/y^{k_2},\dots,z_n/y^{k_n})$,
	we have
    \begin{align*}
        h(Z,Y)
        &= -\int_\RL \int_{\RL^n} f_Y(y) f_{Z|Y}(z|y)\log(f_Y(y)f_{Z}(z|y))
	\,dzdy
        \\ &= -\int_\RL\int_{\RL^n} \frac{1}{\abs{y}^K}f_Y(y)f_{X|Y}
	\left(z'\middle|y\right)\log\left(
	\frac{1}{\abs{y}^K}f_Y(y)f_{X|Y}\left(z'\middle|y\right)\right)\,|y|^Kdz'dy
        \\ &= -\int_\RL \int_{\RL^n} f_Y(y)f_{X|Y}(x|y)\log\left(\frac{1}{\abs{y}^K}f_Y(y)f_{X|Y}(x|y)\right)\,dxdy
        \\ &= h(Y) + h(X|Y) + K\cdot\BBE[\log\abs{Y}],
    \end{align*}
    which rearranges to the desired 
	identity~(\ref{eq:differential_entropy_random_multiplication}).
The second identity follows from~(\ref{eq:entent}).
\qed

The following is the 
multiplicative version 
of Ruzsa's triangle inequality.

\begin{proposition}[Multiplicative Ruzsa triangle inequality]\label{proposition:multiplicative_Ruzsa_triangle}
    Let $X$, $Y$, and $Z$ be independent continuous random variables 
	such that $\log\abs{X}$,  $\log\abs{Y}$, and $\log\abs{Z}$ 
	are integrable.
    Then
    \begin{equation*}
        h(X/Z) \leq h(X/Y) + h(Y/Z) - h(Y) + \BBE[\log\abs{Y}],
    \end{equation*}
    or, equivalently,
    \[ \widetilde h(X/Z) \leq \widetilde h(X/Y) + \widetilde h(Y/Z) - \widetilde h(Y). \]
\end{proposition}

Similarly, the multiplicative analog of
the submodularity-for-sums bound in~(\ref{old_eq:sum_submodularity})
is:

\begin{proposition}\label{proposition:3122_multiplicative}
    Let $X$, $Y$, and $Z$ be independent, continuous random variables, 
	such that $\log\abs{X}$, $\log|Y|$ and $\log|Z|$ are integrable.
    Then
    \[ h(XYZ) + h(Y) \leq h(XY) + h(YZ), \]
    or, equivalently,
    \[ \widetilde h(XYZ) + \widetilde h(Y) \leq \widetilde h(XY) + \widetilde h(YZ). \]
\end{proposition}

Next, we define the {\em multiplicative Ruzsa distance}
between two continuous random variables $X$ and $Y$, 
where both $\log\abs{X}$ and $\log\abs{Y}$ are integrable, 
in analog with the additive version in~(\ref{eq:contRuzsa}),
as,
\begin{align*} 
\widetilde{d}(X,Y) 
&= 
	h(X'/Y')-\frac{1}{2}h(X)-\frac{1}{2}h(1/Y) 
	- \frac{1}{2}\BBE[\log\abs{X}] + \frac{1}{2}\BBE[\log\abs{Y}]\\
&= 
	\widetilde h(X'/Y')-\frac{1}{2}\widetilde h(X) 
	- \frac{1}{2}\widetilde h(Y), 
\end{align*}
where $X'\sim X$ and $Y'\sim Y$ are independent.

Note that the regular Ruzsa distance is translation invariant,
$d(X+c_1,Y+c_2)=d(X,Y)$, while the multiplicative Ruzsa distance
is scale invariant,
$\widetilde{d}(c_1 X, c_2 Y) = \widetilde{d}(X,Y)$.

\subsection{The multiplicative doubling constant}
\label{s:multisX}

The doubling constant of a discrete random variable $X$
was defined in~(\ref{eq:doublingH}) as
$s(X)=H(X+X')-H(X)$, with $X,X'$ being i.i.d.
Similarly, we define the {\em doubling constant}
$\sigma(X)$ and the {\em difference constant}
$\delta(X)$ of a continuous random variable $X$
as,
$$\sigma(X)=h(X+X')-h(X),\qquad \delta(X)=h(X-X')-h(X),
$$
where $X,X'$ are i.i.d.
In this notation, 
the doubling-difference inequality 
for differential entropy described
in Section~\ref{s:diffh},
states that
\bqq \frac{1}{2} \leq \frac{\sigma(X)}{\delta(X)} \leq 2. 
\label{eq:DDh}
\eqq

The natural multiplicative analog of $\sigma(X)$
and $\delta(X)$ are,
\[ \widetilde\sigma(X):=h(XX')-h(X)-\BBE[\log\abs{X}] 
= \widetilde{d}(X,1/X),\]
and,
\[ \widetilde\delta(X)=h(X/X')-h(X)+\BBE[\log\abs{X}] 
= \widetilde{d}(X,X),\]
where $X,X'$ are i.i.d.
The following
``square-quotient inequality''
is the multiplicative version of the doubling-difference inequality.

\begin{theorem}[Square-quotient inequality]
\label{theorem:multiplicative_doubling_difference}
    Let $X$ be a continuous random variable 
such that $\log\abs{X}$ is integrable.
    Then
    \begin{equation*}
        \frac{1}{2} \leq \frac{\widetilde\sigma(X)}{\widetilde\delta(X)}\leq 2,
    \end{equation*}
    or, equivalently,
    \begin{equation*}
        \frac{1}{2}\widetilde{d}(X,X) \leq \widetilde{d}(X,1/X) \leq 2\widetilde{d}(X,X).
    \end{equation*}
\end{theorem}

Finally, we observe that the same
inductive application of Proposition~\ref{proposition:3122_multiplicative}
as was done for the additive case in~\cite{KM:16},
immediately gives the following
multiplicative version
of the Pl\"unnecke--Ruzsa inequality.

\begin{theorem}[Multiplicative Pl\"unnecke--Ruzsa inequality]
\label{theorem:multiplicative_PR}
    Let $X$ and $Y_1,Y_2,\dots,Y_n$ be independent, continuous
random variables, such that $\log\abs{X}$ 
and all $\log|Y_i|$, $1\leq i\leq n$, are integrable.
Suppose there are finite
constants $K_1,K_2,\dots,K_n$ satisfying 
$h(XY_i)\leq h(X)+\log K_i$ for each $i$.
    Then:
    \begin{equation*}
        h(XY_1Y_2\cdots Y_n) \leq h(X) + \log (K_1 K_2 \cdots K_n).
    \end{equation*}
\end{theorem}

\section{Differential entropy bounds for sum-product combinations}
\label{section:sum_product}

The main goal of this section is to establish two general versions
of the Pl\"unnecke--Ruzsa inequality that involve both sums 
and products of i.i.d.\ random variables. The first one,
in Theorem~\ref{theorem:general_ring_PR},
applies to real-valued continuous random variables
and the second one, in Theorem~\ref{theorem:general_ring_PR_discrete},
holds for discrete random
variables taking values in an arbitrary integral domain.

We begin by proving
the following inequality for sums and
products of an arbitrary triple $(X,Y,Z)$
of continuous random variables.
It is the continuous version
of a corresponding inequality proved for discrete
entropy by
M\'ath\'e and O'Regan in~\cite[Proposition~4.1]{mathe:23}.
Proposition~\ref{proposition:modified_MO_Prop4.1} will be
used in the proof of our first version of the 
ring Pl\"unnecke--Ruzsa inequality in
Theorem~\ref{theorem:ring_PR}.

\begin{proposition}\label{proposition:modified_MO_Prop4.1}
    Let $X,Y,Z$ be continuous random variables 
such that $\log\abs{X}$ is integrable.
    Then
    \begin{equation}\label{eq:modified_MO_Prop4.1}
        h(X(Y+Z)) + h(X,Y,Z) \leq h(X,Y+Z) + h(XY,XZ) - \BBE[\log\abs{X}].
    \end{equation}
    In particular, if $X$, $Y$, and $Z$ are i.i.d., then
    \begin{equation}\label{eq:modified_MO_Prop4.1eq2}
        h(X(Y\pm Z)) \leq 2h(XY) + h(X\pm Y) - 2h(X) - \BBE[\log\abs{X}],
    \end{equation}
where the inequality holds with either choice of signs, $(+,+)$
or $(-,-)$.
\end{proposition}

\noindent
{\sc Proof. }
{
Recall that
$h(aW)=h(W)+\log|a|$ for any continuous random variable $W$ 
and any nonzero constant $a$.
Then we note that we may express}
    \begin{align*}
        I(X(Y+Z);X)
        &= h(X(Y+Z)) - h(X(Y+Z)|X)
        \\ &= h(X(Y+Z)) - h(Y+Z|X) - \BBE[\log\abs{X}]
        \\ &= h(X(Y+Z)) - h(X,Y+Z) + h(X) - \BBE[\log\abs{X}],
    \end{align*}
    and
    \begin{align*}
        I(XY,XZ;X)
        &= h(XY,XZ) - h(XY,XZ|X)
        \\ &= h(XY,XZ) - h(Y,Z|X) - 2\BBE[\log\abs{X}]
        \\ &= h(XY,XZ) - h(X,Y,Z) + h(X) - 2\BBE[\log\abs{X}].
    \end{align*}
    By the data processing inequality, we have
    \[ I(X(Y+Z);X) \leq I(XY,XZ;X), \]
    which rearranges to~\eqref{eq:modified_MO_Prop4.1}.

Assuming $X$, $Y$, and $Z$ are i.i.d., 
applying~\eqref{eq:modified_MO_Prop4.1} to $X$, $Y$, and $\pm Z$,
and using the simple fact that
    \begin{equation*}
        h(XY,XZ) \leq h(XY) + h(XZ) = 2h(XY). 
    \end{equation*}
yields~\eqref{eq:modified_MO_Prop4.1eq2}.
\qed

\subsection{Ring Pl\"unnecke--Ruzsa inequality}

We begin by first establishing the following
simple version of a  ``ring'' Pl\"unnecke--Ruzsa 
inequality, which provides an upper bound
to $h(XY-ZW)$ when $X,Y,Z,W$ are i.i.d.
Recall the definitions of the doubling constant
$\sigma(X)$ and of the associated constants
$\delta(X)$ and $\widetilde{\sigma}(X)$ from Section~\ref{s:multisX}.

\begin{theorem}[Ring Pl\"unnecke--Ruzsa]\label{theorem:ring_PR}
    If $X,Y,Z,W$ are i.i.d.\ continuous random variables
	such that $\log\abs{X}$ is integrable, then,
    \begin{align}\label{eq:ring_PR_1}
        h(XY-ZW) &\leq 5h(XY) + 2h(X+Y) - 6h(X) - 4\BBE[\log\abs{X}]
        \\ &= h(XY) + 4\widetilde{\sigma}(X) + 2\sigma(X), \notag
    \end{align}
    and
    \begin{align}\label{eq:ring_PR_2}
        h(XY+ZW) &\leq 5h(XY) + h(X+Y) + h(X-Y) - 6h(X) - 4\BBE[\log\abs{X}]
        \\ &= h(XY) + 4\widetilde{\sigma}(X) + \sigma(X) + \delta(X). \notag
    \end{align}
    In addition,
    \begin{equation}\label{eq:ring_PR_3}
        h(XY+ZW) \leq 5h(XY) + 3h(X\pm Y) - 7h(X) - 4\BBE[\log\abs{X}].
    \end{equation}
\end{theorem}

For the proof we will need the following bound.

\begin{lemma}\label{lemma:helper_ring_PR}
	For any three continuous random variables $X,Y,Z$,
    \begin{equation}\label{eq:helper_ring_PR}
        h(X\pm Y) + h(X,Y,Z) \leq h(X,Y) + h(X\mp Z,Y+Z),
    \end{equation}
	where the inequality holds with either choice of 
	signs,
	$(+, -)$ or $(-, +)$.
\end{lemma}

\noindent
{\sc Proof. }
    Note that
    \begin{align*}
        I(X\pm Y;Y+Z)
        &= h(X\pm Y) + h(Y+Z) - h(X\pm Y,Y+Z)
        \\ &= h(X\pm Y) + h(Y+Z) - h(X\mp Z,Y+Z)
    \end{align*}
    and
    \begin{align*}
        I(X,Y;Y+Z)
        &= h(X,Y) + h(Y+Z) - h(X,Y,Y+Z)
        \\ &= h(X,Y) + h(Y+Z) - h(X,Y,Z).
    \end{align*}
    By the data processing inequality, we have
    \[ I(X\pm Y;Y+Z) \leq I(X,Y;Y+Z), \]
    which rearranges to~\eqref{eq:helper_ring_PR}.
\qed

{\sc Proof of Theorem~\ref{theorem:ring_PR}. }
    Let $V=(XY,ZW,XW)$.
    Note that by Lemma~\ref{lemma:differential_entropy_random_multiplication},
    \begin{align*}
        h(V|X)
        &= h(Y,ZW,W|X) + 2\BBE[\log\abs{X}]
        \\ &= h(Y) + h(ZW,W) + 2\BBE[\log\abs{X}]
        \\ &= 2h(X) + h(ZW|W) + 2\BBE[\log\abs{X}]
        \\ &= 2h(X) + h(Z|W) + 3\BBE[\log\abs{X}]
        \\ &= 3h(X) + 3\BBE[\log\abs{X}].
    \end{align*}
    Also,
    \begin{align*}
        h(X|V)
        &\leq h(X|XY)
        \\ &= h(X,XY) - h(XY)
        \\ &= h(X) + h(Y|X) - h(XY) + \BBE[\log\abs{X}]
        \\ &= 2h(X) - h(XY) + \BBE[\log\abs{X}],
    \end{align*}
    which implies the lower bound
    \begin{align}
        h(V)
        &= h(V|X) + h(X) - h(X|V)
        \nonumber\\ 
	&\geq 2h(X) + h(XY) + 2\BBE[\log\abs{X}].
	\label{eq:PRLB}
    \end{align}
    Applying Lemma~\ref{lemma:helper_ring_PR} to the three 
	random variables $XY,ZW,XW$, we obtain
    \begin{align*}
        h(XY\pm ZW) + h(V) &\leq h(XY,ZW) + h(XY\mp XW,ZW+XW).
    \end{align*}
    This, combined with the lower bound in~(\ref{eq:PRLB}),  yields
    \begin{align*}
        h(XY\pm ZW) + 2h(X) + h(XY) + 2\BBE[\log\abs{X}] 
        &\leq h(XY,ZW) + h(XY\mp XW,ZW+XW)
        \\ &\leq h(XY,ZW) + h(XY\mp XW) + h(ZW+XW),
    \end{align*}
    and using the fact that $X,Y,Z,W$ are i.i.d.\ and rearranging,
    \begin{align*}
        h(XY\pm ZW) \leq h(XY) + h(X(Y+Z)) + h(X(Y\mp Z)) - 2h(X) - 2\BBE[\log\abs{X}].
    \end{align*}
    Finally, applying Proposition~\ref{proposition:modified_MO_Prop4.1} 
	yields~\eqref{eq:ring_PR_1} and~\eqref{eq:ring_PR_2},
	and the doubling-difference inequality~(\ref{eq:DDh})
    combined with~\eqref{eq:ring_PR_2}, yields~\eqref{eq:ring_PR_3}.
\qed

A more general, inductive version of the argument in the last
proof, allows us to prove the following technical result,
which is the key step in establishing 
the general ring Pl\"unnecke--Ruzsa inequality below.

\begin{proposition}\label{proposition:ring_PR_AB}
    Suppose the continuous random variables
	 $X,X_1,X_2,\dots,X_n,Y_1,Y_2,\dots,Y_n$ are i.i.d.,
	 and that $\log\abs{X}$ is integrable.
    Then
    \begin{align*}
        h(X_1X_2\cdots X_n + Y_1Y_2\cdots Y_n)
        &\leq 3h(X_1X_2\cdots X_n) + 2\sum_{k=2}^n h(X_1X_2\cdots X_k) + (n-1)h(X_1-Y_1) 
        \\ &\quad + h(X_1 + Y_1) - 3nh(X) - (n+2)(n-1)\BBE[\log\abs{X}].
    \end{align*}
\end{proposition}

\noindent
{\sc Proof. }
    We prove the result by induction.
    The case $n=1$ is trivial, and the case $n=2$ follows 
	from Theorem~\ref{theorem:ring_PR}.
    For the inductive step, assume $n\geq 3$.
    Let $A=X_1X_2\cdots X_{n-1}$ and let $B=Y_1Y_2\cdots Y_{n-1}$.
    We mimic the proof of Theorem~\ref{theorem:ring_PR}, except using the 
	less symmetric random vector $Z=(AX_n,BY_n,AY_n)$ in place of $V$.
    By Lemma~\ref{lemma:differential_entropy_random_multiplication},
    \begin{align*}
        h(Z|A) 
        &= h(X_n,BY_n,Y_n) + 2(n-1)\BBE[\log\abs{X}]
        \\ &= 2h(X) + h(BY_n|Y_n) + 2(n-1)\BBE[\log\abs{X}]
        \\ &= 2h(X) + h(A) + (2n-1)\BBE[\log\abs{X}].
    \end{align*}
	We also have,
    \begin{align*}
        h(A|Z)
        &\leq h(A|AX_n)
        \\ &= h(A) + h(AX_n|A) - h(AX_n)
        \\ &= h(A) + h(X_n) - h(AX_n) + (n-1)\BBE[\log\abs{X}].
    \end{align*}
    Thus,
    \begin{align}
        h(Z)
        &= h(Z|A) + h(A) - h(A|Z)
        \nonumber\\ 
	&\geq h(AX_n) + h(A) + h(X) + n\BBE[\log\abs{X}].
	\label{eq:ZLB}
    \end{align}
    Meanwhile, using Lemma~\ref{lemma:helper_ring_PR} applied to $Z$,
	and Proposition~\ref{proposition:modified_MO_Prop4.1},
	we obtain,
    \begin{align*}
        h(AX_n+BY_n)+h(Z) 
        &\leq h(AX_n,BY_n) + h(A(X_n-Y_n),(A+B)Y_n)
        \\ &\leq 2h(AX_n) + h(A(X_n-Y_n)) + h((A+B)Y_n)
        \\ &\leq 2h(AX_n) + h(A,X_n-Y_n) - h(A,X_n,Y_n) + h(AX_n,AY_n)
        \\ &\quad -(n-1)\BBE[\log\abs{X}] + h(A+B,Y_n) - h(A,B,Y_n) 
        \\ &\quad + h(AY_n,BY_n) - \BBE[\log\abs{X}]
        \\ &\leq 6h(AX_n) + h(A+B) + h(X_n-Y_n) -2h(X) -2h(A) -n\BBE[\log\abs{X}],
    \end{align*}
    and using the lower bound~(\ref{eq:ZLB}) on $h(Z)$, gives,
    \begin{align*}
        h(AX_n+BY_n)
        &\leq 5h(AX_n) + h(A+B) + h(X_n-Y_n) - 3h(X) - 3h(A) - 2n\BBE[\log\abs{X}].
    \end{align*}
    Finally, applying the inductive hypothesis to $h(A+B)$,
    \begin{align*}
        h(AX_n+BY_n)
        &\leq 5h(AX_n) + 3h(A) + 2\sum_{k=2}^{n-1} h(X_1X_2\cdots X_k) + (n-2)h(X_1-Y_1) + h(X_1+Y_1) \\ &\quad -3(n-1)h(X) - (n+1)(n-2)\BBE[\log\abs{X}] + h(X_n-Y_n) -3h(X) - 3h(A) 
        \\ &\quad - 2n\BBE[\log\abs{X}]
        \\ &= 3h(AX_n) + 2\sum_{k=2}^n h(X_1X_2\cdots X_k) + (n-1)h(X_1-Y_1) + h(X_1+Y_1) - 3nh(X) 
        \\ &\quad - (n+2)(n-1)\BBE[\log\abs{X}],
    \end{align*}
which is exactly the claimed result.
\qed

Applying the multiplicative Pl\"unnecke--Ruzsa 
inequality, Theorem~\ref{theorem:multiplicative_PR}, 
to Proposition~\ref{proposition:ring_PR_AB}, 
immediately yields the following corollary.

\begin{corollary}
\label{corollary:ring_PR_AB}
    Suppose the continuous random variables
	 $X,X_1,X_2,\dots,X_n,Y_1,Y_2,\dots,Y_n$ are i.i.d.,
	 and that $\log\abs{X}$ is integrable.
    Then
    \begin{align}
        h(X_1X_2\cdots X_n + Y_1Y_2\cdots Y_n) 
        &\leq h(X_1X_2\cdots X_n) + (n+2)(n-1)
	\big[h(X_1Y_1)-h(X)-\BBE[\log\abs{X}]\big] \notag
        \\&\quad+ (n-1)\big[h(X_1-Y_1)-h(X)\big] + h(X_1+Y_1)-h(X)
        \label{eq:ring_PR_AB_condensed}
        \\ &= h(X_1X_2\cdots X_n) + (n+2)(n-1)\widetilde{\sigma}(X) + (n-1)\delta(X) + \sigma(X).
	 \quad\quad\notag
    \end{align}
\end{corollary}

Finally, if we apply
the additive Pl\"unnecke--Ruzsa 
inequality~(\ref{eq:PRadd}) of
\cite{KM:14} 
to~\eqref{eq:ring_PR_AB_condensed},
and then use the multiplicative Pl\"unnecke--Ruzsa inequality,
Theorem~\ref{theorem:multiplicative_PR}, 
we obtain the following general 
ring Pl\"unnecke--Ruzsa inequality.
This is the main result of this section.

\begin{theorem}[General ring Pl\"unnecke--Ruzsa inequality]
\label{theorem:general_ring_PR}
Suppose that continuous random variables
$\{X_{ij}\;;\;1\leq i\leq m,1\leq j\leq n\}$
are i.i.d.\ and distributed as $X$.
Assuming $\log\abs{X}$ is integrable, we have
\begin{align*}
h\left(\sum_{i=1}^m\prod_{j=1}^n X_{i,j}\right)
&\leq
	h(X_{1,1}X_{1,2}\cdots X_{1,n}) 
	+ (m-1)\big[(n+2)(n-1)\widetilde{\sigma}(X)+(n-1)\delta(X) 
        + \sigma(X)\big]\\
&\leq 
	h(X) 
	+ \big((m-1)(n+2)+1\big)(n-1)\widetilde{\sigma}(X) + (m-1)(n-1)\delta(X)
        + (m-1)\sigma(X)\\
&\qquad 
	+ (n-1)\BBE[\log\abs{X}].
    \end{align*}
\end{theorem}

Essentially the same result holds for discrete random variables,
and with the same proof. 
Ignoring the $\BBE[\log|X|]$ terms
and writing
$\sigma(X)=H(X_1+X_2)-H(X)$,
$\widetilde{\sigma}(X)=H(X_1X_2)-H(X)$,
and $\delta(X)=H(X_1-X_2)-H(X)$,
we note the following:
The discrete analogs 
of Proposition~\ref{proposition:modified_MO_Prop4.1} 
and Theorem~\ref{theorem:ring_PR} were proven 
in~\cite{mathe:23}, Lemma~\ref{lemma:helper_ring_PR} holds via the 
same proof for the discrete case, 
Lemmas~\ref{lemma:differential_entropy_inverse}
and~\ref{lemma:differential_entropy_random_multiplication} 
are trivial for discrete entropy,
and Theorem~\ref{theorem:multiplicative_PR} is equivalent to 
the discrete analog of~\cite[Theorem~3.11]{KM:14}, 
which holds via the same proof.
Therefore, we readily obtain:

\begin{theorem}%
[General ring Pl\"unnecke--Ruzsa inequality for discrete entropy]
\label{theorem:general_ring_PR_discrete}
Suppose the discrete random variables
$\{X_{ij}\;;\;1\leq i\leq m,1\leq j\leq n\}$
are i.i.d., taking values in an integral domain,
and distributed as $X$.
Assuming $X\neq 0$ a.s., we have
\begin{align*}
H\left(\sum_{i=1}^m\prod_{j=1}^n X_{i,j}\right) 
&\leq 
	H(X_{1,1}X_{1,2}\cdots X_{1,n}) + (m-1)
	\big[(n+2)(n-1)\widetilde{\sigma}(X)+(n-1)\delta(X) 
        + \sigma(X)\big]\\
&\leq 
	H(X) + \big((m-1)(n+2)+1\big)(n-1)\widetilde{\sigma}(X) 
	+ (m-1)(n-1)\delta(X)
       + (m-1)\sigma(X).
    \end{align*}
\end{theorem}

In the special case when both
$\sigma(X)$ and $\widetilde{\sigma}(X)$ are bounded
by $\log K$, by the doubling-difference inequality
we also have that $\delta(X)\leq 2\log K$,
and the discrete version of our 
general ring Pl\"unnecke--Ruzsa inequality 
implies the bound
$$H\left(\sum_{i=1}^m\prod_{j=1}^n X_{i,j}\right) 
\leq 
	H(X) + \big[(n-1)+(m-1)\big(n^2+3n-3\big)\big]\log K.
$$
Interestingly, this is identical to a result
independently obtained by M\'ath\'e and O'Regan
in an updated arXiv preprint 
version of~\cite{mathe:23}.

\subsection{An inequality for slopes}

Here we prove an inequality for the entropy 
for the ratio of sums (or differences) of i.i.d.\
random variables. It is a partial 
analog of the discrete entropy bound~\cite[Theorem~4.7]{mathe:23}.

\begin{theorem}\label{theorem:slopes}
 Suppose $X,Y,Z,W$ are i.i.d.\ continuous random variables,
such that $\log\abs{X}$ and $\log\abs{X\pm Y}$ are integrable.
    Then
    \[ h\left(\frac{X\pm Y}{Z\pm W}\right) + 5h(X) \leq 4h(XY) + 2h(X\pm Y) - 3\BBE[\log\abs{X}]-2\BBE[\log\abs{X\pm Y}], \]
where the inequality holds with either choice of all-plus
or all-minus signs, under the corresponding assumption.
Equivalently, we have,
    \[ \widetilde h\left(\frac{X\pm Y}{Z\pm W}\right) + 5\widetilde h(X) \leq 4\widetilde h(XY) + 2\widetilde h(X\pm Y). \]
\end{theorem}

Theorem~\ref{theorem:slopes} immediately follows from the 
next lemma, upon
taking all five random variables to be i.i.d.

\begin{lemma}\label{lemma:slopes}
    Suppose $X,Y,Z,W$ and $U$ are independent continuous 
	random variables, such that $\log\abs{U}$,
	 $\log\abs{X\pm Y}$ 
	 and $\log\abs{Z\pm W}$ 
	are integrable.
    Then
    \begin{align*}
        h\left(\frac{X\pm Y}{Z\pm W}\right) + h(X,Y,Z,W,U) 
        &\leq h(X\pm Y) + h(Z\pm W) + h(UX,UY) + h(UZ,UW) 
        \\ &\quad -3\BBE[\log\abs{U}]-2\BBE[\log\abs{Z\pm W}],
    \end{align*}
where the inequality holds with either choice of all-plus
or all-minus signs, under the corresponding pair of assumptions.
Equivalently, we have,
    \[ \widetilde h\left(\frac{X\pm Y}{Z\pm W}\right) + \widetilde h(X,Y,Z,W,U)
\leq \widetilde h(X\pm Y) + \widetilde h(Z\pm W) + \widetilde h(UX,UY) + \widetilde h(UZ,UW). \]
\end{lemma}

\noindent
{\sc Proof. }
    Let $V_1=X\pm Y$ and $V_2=Z\pm W$.
    By Proposition~\ref{proposition:multiplicative_Ruzsa_triangle} applied
to $V_1$, $1/U$, and $V_2$, and Lemma~\ref{lemma:differential_entropy_inverse},
    \begin{align*}
        h(V_1/V_2) + h(U) - 2\BBE[\log\abs{U}] \leq h(V_1U) + h(UV_2) - 3\BBE[\log\abs{U}]-2\BBE[\log\abs{V_2}].
    \end{align*}
    Rearranging and applying Proposition~\ref{proposition:modified_MO_Prop4.1} 
	to $h(V_1U)$ and $h(UV_2)$, we find
    \begin{align*}
        h\left(\frac{X\pm Y}{Z\pm W}\right) + h(U)
        &\leq h(U,X\pm Y) + h(U,Z\pm W) + h(UX,UY) + h(UZ,UW) 
        \\ &\quad - h(U,X,Y) - h(U,Z,W) 
        - 3\BBE[\log\abs{U}]-2\BBE[\log\abs{Z\pm W}].
    \end{align*}
    Using the independence assumption and rearranging 
yields the desired inequality.
\qed

\section{On the entropic sum-product phenomenon in $\IN$}
\label{section:entropic_sumproduct}

For a subset $A$ of the integers $\IN$, 
let $A\cdot A$ denote the product set
$A\cdot A=\{a_1a_2:a_1,a_2\in A\}$.
In 1983, Erd\H os and Szemer\'edi~\cite{erdos:83}
observed 
that $A+A$ and $A\cdot A$ cannot both be small simultaneously.
Specifically, they showed that there is a positive constant
$\varepsilon$ such that, for $A\subset \IN$,
\bqq
\max\set{\abs{A+A},\abs{A\cdot A}} \geq \abs{A}^{1+\varepsilon-o(1)},
\label{eq:E-Z}
\eqq
where the $o(1)$ term tends to zero as $|A|\to\infty$.
Although they did not provide an explicit estimate for $\varepsilon$,
it is widely conjectured -- though still unproven -- that the
result should hold for any $\varepsilon\leq 1$.
The best result to date is by Bloom~\cite{bloom:25},
with $\varepsilon=\frac{1}{3}+\frac{2}{951}$; this result also holds more generally for finite $A\subset\RL$.

In view of the entropy/cardinality correspondence
discussed in the Introduction,
it might be tempting to conjecture an analogous result
for the entropy.
Namely, that for
i.i.d.\ $\IN$-valued random variables $X,X'$,
$$\max\{H(X+X'),H(X\cdot X')\}\geq (2-o(1))H(X),$$
as $H(X)\to\infty$, or at least that,
{as conjectured by Goh~\cite{goh:24},}
\bqq
\max\{H(X+X'),H(X\cdot X')\}\geq (1+\epsilon-o(1))H(X),
\label{eq:entropic_sumproduct}
\eqq
for some $\varepsilon>0$. 
In this section
we show that,
if such a result were to hold, it would necessarily
be with $\varepsilon\leq 1/3$. In particular,
the obvious analog of Bloom's bound
fails for the entropy.

\begin{example}
For a large positive integer $n$ that we will send to infinity, consider the discrete real-valued random variable $X$ which takes value 0 with probability $\frac{1}{3}$, and with probability $\frac{2}{3}$ takes a value uniformly at random in $\{1,2,\dots,n\}$.
Asymptotically as $n\to\infty$, its entropy is $H(X)=\frac{2}{3}\log n+O(1)$.
Let $X'$ be an independent copy of $X$, and let $A$ and $A'$ be the indicators for $X=0$ and $X'=0$, respectively.
Note that $H(X+X'|A=0,A'=0)=\log n+O(1)$, because
\[ H(X+X'|A=0,A'=0) \geq H(X+X'|A=0,A'=0,X') = H(X|A=0) = \log n \]
and
\[ H(X+X'|A=0,A'=0) \leq \log\abs{\set{1,2,\dots,n}+\set{1,2,\dots,n}} = \log(2n-1) \leq \log n+\log 2. \]
Thus,
\[ H(X+X') = H(X+X'|A,A') + O(1) = \paren{1-\frac{1}{9}}\log n+O(1) 
= \frac{8}{9}\log n+O(1), \]
where the first equality follows from the inequalities
\begin{align*}
    H(X+X'|A,A') \leq H(X+X') &\leq H(X+X',A,A')
    \\ &= H(X+X'|A,A') + H(A,A')
    \\ &\leq H(X+X'|A,A') + \log 4.
\end{align*}
We also observe
\[ H(X\cdot X'|A=0,A'=0)\leq \log\abs{\set{1,2,\dots,n}\cdot\set{1,2,\dots,n}}\leq\log(n^2), \]
and so
\[ H(X\cdot X') = H(X\cdot X'|A,A')+O(1) \leq \frac{4}{9}\log(n^2) + O(1) = \frac{8}{9}\log n + O(1), \]
so
\[ \max\set{H(X+X'),H(X\cdot X')}\leq \paren{\frac{4}{3}+o(1)}H(X). \]
By taking $n\to\infty$, we can obtain arbitrarily large $H(X)$, 
so this example shows~\eqref{eq:entropic_sumproduct} cannot hold 
in general for any $\varepsilon>\frac{1}{3}$.
\end{example}

{Note that the choice of the value 1/3 for
$\BBP(X=0)$ is optimal in that any other value gives
a constant larger than $4/3$ in the final result.
Therefore, this method cannot improve our estimate that,
if~(\ref{eq:entropic_sumproduct})
were to hold, it could only hold for some
$\varepsilon\leq 1/3$.}

However, we know~\eqref{eq:E-Z} holds for 
some $\varepsilon>\frac{1}{3}$, and one may wonder whether we 
can use this inequality to show~\eqref{eq:entropic_sumproduct} at least 
in the case when $X$ is uniformly distributed over 
some finite set $A\subset\RL$.
To that end, one may wonder whether there are
$\varepsilon,\varepsilon'\in(0,1)$ such that,
for any finite set $A\subset\RL$, 
$$\abs{A+A}\geq\abs{A}^{1+\varepsilon}
\quad\implies\quad
H(U_A+U_A')\geq(1+\varepsilon'-o(1))H(U_A),$$
where $U_A$ 
and $U_A'$ are i.i.d.\ random variables uniformly distributed on $A$.
The weakest version of this result would be of the form,
``$\abs{A+A}\geq\abs{A}^{2-\varepsilon}$ implies $H(U_A+U_A')\geq(1+\varepsilon-o(1))H(U_A)$, for some small $\varepsilon>0$''.
The following example, a slight generalization of one 
in~\cite{tao:10}, shows that this does not hold for any $\varepsilon>0$.

\begin{example}
Let $n$ be a large positive integer and $\varepsilon>0$.
Take $A$ be the union of $\set{1,2,\dots,n}$ and a set
$B$ consisting of $n^{1-\varepsilon/2}$ 
other integers in general position, i.e., such that
there are no nontrivial pairwise sum relations among
themselves and with $\set{1,2,\dots,n}$:
If $a+b=c+d$ for $a,c,d\in A$ and $b\in B$, then
$\{a,b\}=\{c,d\}$.

For large $n$ we have $\abs{A}\sim n$, 
while,
\[ \abs{A+A}\geq n^{2-\varepsilon/2}\geq\abs{A}^{2-\varepsilon},\]
because of the sums between $\set{1,2,\dots,n}$ and the 
$n^{1-\varepsilon/2}$ integers in $B$.
Let $U_A$ and $U_A'$ be independent and uniformly distributed on $A$,
and write
$Z$ and $Z'$ for the indicators 
of the events, $\{U_A\in\set{1,2,\dots,n}\}$ and 
$\{U_A'\in\set{1,2,\dots,n}\}$, respectively.
Note that $|A+A|\leq |A|^2$ and 
$\BBP(Z=1),\BBP(Z'=1)\to1$ as $n\to\infty$.
Then, $H(U_A+U_A') \sim H(U_A+U_A'|Z,Z')$,
where
$$H(U_A+U_A'|Z,Z') 
    \leq \BBP(Z=1,Z'=1)(\log n +O(1)) + (1-\BBP(Z=1,Z'=1))\log\abs{A+A}
    \sim \log n.$$
Therefore, despite having $\abs{A+A}\geq\abs{A}^{2-\varepsilon}$, 
we still have 
$H(U_A+U_A')<(1+\varepsilon-o(1))H(U_A)$.
\end{example}

%

\bibliography{ik}
 
\end{document}